\documentstyle[12pt,epsfig]{article}

\newcommand{\be}{\begin{equation}}
\newcommand{\ee}{\end{equation}}

\newcommand{\bbf}{\bf}
\newcommand{\ssl}{\sl}

\newcommand{\T}{\mbox{\bf T}}

\newcommand{\ui}{\bf i}
\newcommand{\uj}{\bf j}
\newcommand{\uk}{\bf k}
\newcommand{\ua}{\bf a}
\newcommand{\ub}{\bf b}
\newcommand{\uc}{\bf c}
\newcommand{\mui}{\tau}
\newcommand{\bea}{\begin{eqnarray}}
\newcommand{\eea}{\end{eqnarray}}
\newcommand{\1}{1\!{\rm l}}

\begin{document}
\bibliographystyle{unsrt}

\begin{titlepage}
\begin{flushright}
FSUJ-TPI-99/10 \\
hep-th/9910116  \\
\end{flushright}

\vspace{1 cm}

\begin{center}
\Huge{Euclidean Freedman--Schwarz model}

\vskip12mm
\large
Mikhail S. Volkov\footnote{Supported by 
the Deutsche Forschungsgemeinschaft,
DFG-Wi 777/4-1}

\vspace{3mm}
{\small\sl
Institute for Theoretical Physics\\
Friedrich Schiller University of Jena\\
Max-Wien Platz 1, D-07743\\
Jena, Germany\\
e-mail: vol@tpi.uni-jena.de\\
}

\end{center}

\vspace{10 mm}
\noindent
{The N=4 gauged SU(2)$\times$SU(1,1)
supergravity in four-dimensional Euclidean space
is obtained via a consistent dimensional reduction
of the N=1, D=10 supergravity on 
$S^3\times AdS_3$. 
The dilaton potential in the theory is proportional
to the difference of the two gauge coupling constants,
which is due to the opposite signs of the curvatures of $S^3$ and $AdS_3$.
As a result, the potential 
can be positive, negative, or zero -- depending on the values of the 
constants. A consistent reduction of the fermion supersymmetry
transformations is performed 
at the linearized level, and  special attention is paid
to the Euclidean Majorana condition. 
A further reduction of the D=4 theory is considered  
to the static, purely magnetic sector, 
where the vacuum solutions are studied. 
The Bogomol'nyi equations 
are derived and their essentially 
non-Abelian monopole-type and sphaleron-type
solutions are presented. Any
solution  in the theory
can be uplifted to become
a vacuum of string  or M-theory.
}

\end{titlepage}

\newpage

\section{Introduction}

Supergravity backgrounds play an important role in the analysis
of string theory. Besides genuine fully supersymmetric string vacua,
also solutions with partial supersymmetry
(p-branes, monopoles etc. \cite{Duff95})  are
presently obtaining much consideration, in particular in view
of their role in verifying various duality conjectures. 
Unfortunately, apart from stringy monopoles \cite{Gauntlett93} 
and related solutions \cite{Duff95}
obtained via the heterotic five-brane construction \cite{Strominger90}
(see also \cite{Gibbons94a,Gibbons95a}), most of the literature
is devoted to solutions with Abelian gauge fields. 
This is easily understood, since such configurations can often be obtained
from the known solutions of the Einstein-Maxwell system,
while the non-Abelian sector is much more difficult to study.
On the other hand, it is to be expected that also configurations with
non-Abelian gauge fields will eventually play an important role.
Apart from this, gauged supergravity models obtainable from  string 
or M-theory via the Kaluza-Klein reduction
 have recently regained 
considerable interest in view of the AdS/CFT correspondence;
see for example \cite{Nastase99,Cvetic99,Lu99}.
This also suggests studying classical solutions
of supergravities with non-Abelian gauge fields.
Finally, systems with gravitating Yang-Mills fields
can be studied in the context of General Relativity, where  
they have recently attracted a lot of  attention in view of the 
unusual properties of their solutions  \cite{Volkov98}. 
Unfortunately, due to the
high complexity of the equations, our knowledge is largely based on 
numerical analysis. At the same time, 
gauged supergravities provide the rare opportunity
to obtain analytical solutions via solving the Bogomol'nyi 
equations.

The present work was  motivated by the desire to analytically obtain
certain particle-like solutions for gravitating Yang-Mills fields, which 
requires to identify the corresponding gauged supergravity model.
 In the recent work \cite{Chamseddine97,Chamseddine98} 
non-Abelian partially supersymmetric vacua were obtained
within the N=4 gauged SU(2)$\times$SU(2)  supergravity,
also known as Freedman--Schwarz (FS) model \cite{Freedman78}.
These solutions are globally regular,
but 
not asymptotically flat  -- due to the 
presence of the dilaton potential, 
\be                                                  \label{0}
{\rm U}(\phi)=-\frac18\,(g_{(1)}^2+g_{(2)}^2)\, {\rm e}^{-2\phi}\, ,
\ee
where $g_{(1)}$ and $g_{(2)}$ are the  gauge coupling constants. 
It is well-known that a dilaton potential
unbounded from below  is generically present
in gauged supergravities (see, however,
Ref.\cite{Gunaydin84}).  The potential 
renders solutions non-asymptotically flat,
and it is therefore necessary to get rid of it if one wants to 
obtain particle-like solutions.
For this purpose the following
trick was employed in \cite{Volkov99}: to truncate
the FS model to the purely magnetic sector and then to pass to 
 imaginary values of the gauge coupling constant $g_{(2)}$:
\be                                                    \label{0a}
g_{(2)}\to i g_{(2)}.
\ee
For $|g_{(2)}|=g_{(1)}$ the potential vanishes. 
Surprisingly, such a formal trick
does not destroy supersymmetry -- the replacement 
(\ref{0a}) in the FS fermion supersymmetry transformations
leads to non-trivial Bogomol'nyi equations. These admit
asymptotically flat solutions.  
As the existence of Bogomol'nyi equations is usually related
to supersymmetry,  it was conjectured
in \cite{Volkov99} that there is another, hitherto unknown  
consistent 
gauged supergravity that can be formally
related to the FS model via the replacement (\ref{0a}). 
The justification of this conjecture is the main subject
of the present paper. 

To understand what the new supergravity is, let us remember that
 the FS model can be obtained via dimensional reduction 
of  the N=1, D=10 supergravity on the  
group manifold SU(2)$\times$SU(2)
\cite{Chamseddine98}. 
Now, the replacement (\ref{0a}) 
suggests considering another 
reduction of the same theory: 
on the  group manifold SU(2)$\times$SU(1,1). 
Since SU(1,1) is non-compact and its
invariant metric is non-positive definite, 
the timelike coordinate of the ten-dimensional space  should  be 
viewed as one of the internal coordinates. Specifically, in order
to match the metric signature  in ten dimensions, one chooses 
a positive Cartan metric for the SU(2) factor and a {\it negative}
one for the SU(1,1) factor. The geometry on the internal
six-space is then described by the standard metric on 
$S^3\times AdS_3$ with the signature $(+++++-)$, while
the remaining four-space becomes Euclidean. 

The FS dilaton potential (\ref{0}) arises upon 
reduction as the contribution of the scalar curvatures
of the internal manifolds (and also due to the D=10 three-form). 
Since the scalar curvatures for the two $S^3$ factors are positive,
the radii of the spheres being $1/g_{(1)}$ and $1/g_{(2)}$, 
the result is proportional to the sum $g_{(1)}^2+g_{(2)}^2$. 
Now, in the case of the reduction on $S^3\times AdS_3$ 
the scalar curvatures for these two factors have different
signs. As a result, choosing again $1/g_{(1)}$ and $1/g_{(2)}$
to be the (real) radii of the internal manifolds, 
the dilaton potential of the resulting 
Euclidean theory is proportional to the {difference} 
$g_{(1)}^2-g_{(2)}^2$. 

To summarize, the new theory appears to be 
an N=4 gauged supergravity in four-dimensional Euclidean space. 
We call it Euclidean Freed\-man-Schwarz (EFS) model.
Its matter content is similar to the one of the FS model, 
but the gauge group is now SU(2)$\times$SU(1,1). The 
dilaton potential 
can be positive,
negative or zero -- depending on the values of the two gauge coupling
constants. This allows one to study various supersymmetric solutions
for gravitating Yang-Mills fields by integrating the 
first order Bogomol'nyi
equations.

The above qualitative considerations will be confirmed by detailed
calculations. In Sec.2 we describe the dimensional 
reduction procedure in the bosonic sector. We use the general
recipes for the reduction on group manifolds given in \cite{Scherk79}
and, in order to keep control over the calculations at every step, 
consider 
the reductions on $S^3\times AdS_3$ and $S^3\times S^3$
simultaneously. Specifically,  
our dimensional reduction ansatz and most of  
other formulas contain a selective parameter, $s$. 
The value $s=1$
corresponds to the reduction on $S^3\times AdS_3$, while
for $s=\sqrt{-1}$ we recover the results of the analysis for the 
$S^3\times S^3$ case described in \cite{Chamseddine98}. 
Since we are interested in obtaining a specific 
rather than the most general four-dimensional model, we
truncate many of the degrees of freedom
and always work at the level of equations of motion
in order to maintain consistency. 
The main result of Sec.2 is that 
a consistent reduction on $S^3\times AdS_3$ is possible and 
the equations of motion of the 
resulting  theory in D=4 can be obtained by varying 
the  Lagrangian in Eq. (\ref{c26}). 
If a four-dimensional configuration fulfills these equations
then its uplifted ten-dimensional version will be on shell. 

In Sec.3 we consider the reduction of the D=10 linearized fermion 
supersymmetry transformations down to D=4. 
After discussing the Euclidean Majorana condition we 
derive the four-dimensional supersymmetry variations
in Eqs. (\ref{d24}), (\ref{d32}).  If these variations vanish for 
a given four-dimensional configuration, then its uplifted version
will be supersymmetric in the ten-dimensional sense. 
This completes the dimensional reduction procedure, as we 
obtain the bosonic Lagrangian and the fermion supersymmetry
transformations, which is sufficient for deriving the Bogomol'nyi
equations. 

In Sec.4 we apply our results in order to obtain supersymmetric
vacua with gravitating Yang-Mills fields by deriving
and integrating the Bogomol'nyi equations. 
First, we consider a further truncation of the theory to the 
static, purely magnetic sector. It turns out that, in this sector, 
the field equations and supersymmetry 
transformations for the FS and EFS models are formally
related via the `analytic continuation' (\ref{0a}). This 
provides a complete explanation of the conjecture of Ref.\cite{Volkov99}. 
We then impose a spherical symmetry and  derive
the supersymmetry constraints, whose consistency 
conditions give us a system of first order non-linear 
Bogomol'nyi equations. Essentially non-Abelian
solutions of these equations are known in two special cases 
\cite{Chamseddine97,Chamseddine98,Volkov99}. 
For $g_{(1)}\neq 0$, $g_{(2)}=0$ 
the solution preserves 1/4 of the supersymmetries and turns out to 
be of regular monopole type. For $g_{(1)}=g_{(2)}\neq 0$ the solution
is of sphaleron type and 
has only 1/8 of the supersymmetries unbroken. 

The last section contains some concluding remarks. 
We use  units where $\hbar=c=4\pi G=1$.

\section{Bosons}
\setcounter{equation}{0}

We start from the bosonic part of the action of D=10, N=1 supergravity
\begin{equation}
S_{10}=\int \left(
\frac{1}{4}\,\hat{R}
-\frac{1}{2}\,\partial_{M}\hat{\phi}\,\partial ^{M}\hat{\phi}
-\frac{1}{12}\,e^{-2\hat{\phi}}\,
\hat{H}_{MNP}\,\hat{H}^{MNP}\right)
\sqrt{-\hat{g}}\,d^{10}\hat{x},                           \label{c0}
\end{equation}
whose equations of motion are 
\bea                                                            \label{c1}
\hat{\nabla}_M\hat{\nabla}^M\, \hat{\phi}=
-\frac{1}{6}\, e^{-2\hat{\phi}}\, \hat{H}_{MNP}\hat{H}^{MNP}\, , \\
\hat{\nabla}_M\left(e^{-2\hat{\phi}}\, \hat{H}^{MNP}\right)=0\, 
,\label{c2}\\
\hat{R}_{MN}=2\, \partial_M\hat{\phi}\, \partial_N\hat{\phi}
+e^{-2\hat{\phi}}\, \hat{H}_{MPQ}\hat{H}_{N}^{\ \  PQ}
-\frac{1}{12}\, e^{-2\hat{\phi}}\, \hat{g}_{MN}\,
 \hat{H}_{PQS}\hat{H}^{PQS}.                                       \label{c3}
\eea
Our notation is as follows. The hatted symbols are reserved for 
D=10 quantities. We shall always use late letters
for base space indices  and early ones for tangent space indices. 
Indices in ten, four, and six dimensions are 
denoted by capital Latin, small Greek, and small Latin letters, respectively,
such that 
$M\equiv(\mu =0,\ldots ,3;\,{\rm m}=1,\ldots ,6)$ is the base space index,
and 
$A\equiv(\alpha =0,\ldots,3;\,{\rm a}=1,\ldots ,6)$ 
is the tangent space index. 
The spacetime coordinates are $\hat{x}^{M}\equiv (x^{{\mu }},z^{\rm m})$.
We shall sometimes be considering the further split of the four-indices
into 3+1 as $\mu\equiv(0,\uk)$ and $\alpha\equiv(0,\ua)$ 
with the three-indices
denoted by bold-faced letters.  
The 6--space will be assumed to be 
a direct product of two three-dimensional group spaces labeled 
by $(\sigma)=1,2$ with the indices for each 
of the internal three-spaces denoted by Italic letters. 
As a result, every 6-index will be replaced by a pair 
of indices as
${\rm m}\equiv (({\sigma}),i)$ and ${\rm a}\equiv ((\sigma),a)$,
such that $z^{\rm m}\equiv z^{(\sigma)i}$, say. Unless explicitly stated,
we do not assume summation over  repeated indices $(\sigma)$.

The D=10 metric is related to the vielbein,
$\hat{{g}}_{MN}=\hat{\eta}_{AB}\,
\hat{\Theta}_{\ M}^{A}\hat{\Theta}_{\ N}^{B}$,
where 
\be                                                             \label{c6a}
\hat{\eta}_{AB}={\rm diag}
(\overbrace{\underbrace{s^2,+1,+1,+1}_{\rm 4-space}}^{\eta_{\alpha\beta}}
,\underbrace{\overbrace{+1,+1,+1}^{\eta^{(1)}_{ab}}
,\overbrace{+1,+1,-s^2}^{\eta^{(2)}_{ab}}}_{6-{\rm space}}).
\ee
Here the parameter $s$ assumes two values: $s=1$ or $s=i=\sqrt{-1}$.
The two options correspond
to the same theory in D=10 -- up to a renumbering of coordinates --
but to two different choices of the four-space. 
For $s=i$ the time coordinate in D=10 is $\hat{x}^0$
and the four-metric is Lorentzian. 
For $s=1$ the time is $\hat{x}^9$, which is
regarded as one of the internal coordinates, and the four-space is Euclidean. 

The invariant 1-forms on the six-space are denoted by 
$\tilde{\theta}^{\rm a}$, 
the invariant vectors being $\tilde{e}_{\rm b}$, one has
$\langle\tilde{\theta}^{\rm a},\tilde{e}_{\rm b}\rangle
=\delta^{\rm a}_{\rm b}$.
The tetrad vectors and dual 1-forms for the
four-space are $e_\alpha$ and $\theta^\beta$, the four-metric being 
$g_{\mu\nu}=\eta_{\alpha\beta}\theta^\alpha_{\ \mu}\theta^\beta_{\ \nu}$ 
with $\eta_{\alpha\beta}$ defined in (\ref{c6a}). 
The D=10 vielbein vectors are 
$\hat{E}_A$, they are dual to the 1-forms $\hat{\Theta}^B$. 
Our sign conventions for the Riemann and Ricci tensors are
$\hat{R}^{P}_{\ QMN}=
\partial_{M}\hat{\Gamma}^{P}_{QN}-\ldots\, $
and $\hat{R}_{MN}=\hat{R}^{Q}_{\ MQN}$.

\subsection{The dimensional reduction ansatz}
Our goal is to find a parameterization of $\hat{g}_{MN}$, 
$\hat{H}_{MNP}$ and $\hat{\phi}$ in terms of 
four-dimensional variables which reduces Eqs. (\ref{c1})--(\ref{c3})
to a consistent system of 
four-dimensional equations.  As a first step,
we choose $\hat{g}_{MN}$ as
\be                                          \label{c7}
d\hat{s}^2
={\rm e}^{-3\phi/2}\,g_{\mu\nu}dx^\mu dx^\nu
+{\rm e}^{\phi/2}\gamma_{\rm ab}
(A^{\rm a}_{\mu}dx^\mu-\tilde{\theta}^{\rm a}_{\rm m} dz^{\rm m})
(A^{\rm b}_{\nu}dx^\nu-\tilde{\theta}^{\rm b}_{\rm n} dz^{\rm n}).
\ee
Here  the four-metric $g_{\mu\nu}$, the four-dilaton
$\phi=-\frac12\,\hat{\phi}$,
and the fields $A^{\rm a}_{\mu}$ depend only on $x^\mu$,
while the one-forms
$\tilde{\theta}^{\rm a}\equiv \tilde{\theta}^{\rm a}_{\rm m}\, dz^{\rm m}$
are functions of only the internal coordinates $z^{\rm m}$. 
The matrix $\gamma_{\rm ab}$ is assumed to be constant and diagonal.
We assume also that $z^{\rm m}$ span 
a semi-simple (and not necessarily compact) group space ${\cal G}$,
and that  
$\tilde{\theta}^{\rm a}$ are the invariant forms.
The vectors $\tilde{e}_{\rm a}$
dual to $\tilde{\theta}^{\rm a}$ 
satisfy the commutation relations
\be                                                  \label{c8}
[\tilde{e}_{\rm a},\tilde{e}_{\rm b}]=
f_{\rm\ \, ab}^{\rm c}\,\tilde{e}_{\rm c}\, ,
\ee
with $f_{\rm\ ab}^{\rm c}$ being 
the structure constants of ${\cal G}$ .
As a result, we can 
view the metric coefficients $A^{\rm a}_{\mu}$  as
a four-dimensional Yang-Mills
field for the gauge group ${\cal G}$. The
gauge field tensor is
\be                                                              \label{c8a}
F^{\rm a}_{\ \mu\nu}=
\partial_{\mu}A^{\rm a}_{\nu}-\partial_{\nu}A^{\rm a}_{\mu}
+f_{\rm\ \, bc}^{\rm a}A^{\rm b}_{\mu}A^{\rm c}_{\nu}\, .
\ee
We shall assume the group ${\cal G}$ to be the direct product:
\be
{\cal G}={\cal G}^{(1)}\otimes {\cal G}^{(2)}\, .
\ee
Here ${\cal G}^{(1)}$=SU(2), while there are two options
for the second factor 
in the product, ${\cal G}^{(2)}\equiv{\cal G}^{(2)}_s$, where
${\cal G}^{(2)}_s$=SU(1,1) for $s=1$ and  
${\cal G}^{(2)}_s$=SU(2) for $s=i$.  
The direct product structure  implies 
that the invariant group metric, structure constants, etc., decompose into
direct sums. For example, 
\be                                                              \label{c11}
f_{\rm\ \, ab}^{\rm c}=f_{\ \ \ \ ab}^{(1)c}\oplus f_{\ \ \ \ ab}^{(2)c}\, ,
\ee
where
\be                                                             \label{c11a}
f_{\ \ \ \ ab}^{(\sigma)c}=\eta^{(\sigma)cd}\,\varepsilon_{dab}\, ,
\ee
with  $\eta^{(\sigma)}_{cq}$ ($\sigma=1,2$) defined in (\ref{c6a}) and
$\varepsilon_{abc}$ being the antisymmetric tensor ($\varepsilon_{123}=1$).
In addition, we assume that all other quantities
that carry internal indices
also split into direct sums. For example, 
\be                                                       \label{c12}
A^{\rm a}_{\mu}=A^{(1)a}_{\mu}\oplus A^{(2)a}_{\mu},\ \ \ \ \
F^{\rm a}_{\mu\nu}=F^{(1) a}_{\mu\nu}\oplus F^{(2) a}_{\mu\nu}\, ,
\ee
where 
\be                                                       \label{c13}
F^{(\sigma) a}_{\mu\nu}=
\partial_{\mu}A^{(\sigma) a}_{\nu}-\partial_{\nu}A^{(\sigma) a}_{\mu}
+f_{\ \ \ \ bc}^{(\sigma) a}A^{(\sigma) b}_{\mu}A^{(\sigma) c}_{\nu}.
\ee
In particular, we choose 
\be                                                   \label{c9g}
\gamma_{\rm ab}=
\frac{2}{g_{(1)}^{2}}\,\eta^{(1)}_{ab}\oplus
\frac{2}{g_{(2)}^{2}}\,\eta^{(2)}_{ab}\, ,
\ee
with $g_{(1)}$ and $g_{(2)}$ being real constants. 
As a result, we can replace 
the six-dimensional indices 
in all formulas 
by the three-dimensional ones  at the expense of
adding the index $(\sigma)=1,2$. For example, we shall often write 
$\gamma_{ab}^{(\sigma)}$ instead of $\gamma_{\rm ab}$. 

Each of the two factors in  (\ref{c9g})
is proportional to the Cartan metric for the corresponding
group space. We note that for $s=1$
the proportionality coefficients have 
different signs. Specifically, normalizing the Cartan metric as
\be                                                     \label{c9a}
K^{(\sigma)}_{ab}=-\frac12\,
f^{(\sigma)c}_{\ \ \ \ da}f^{(\sigma)d}_{\ \ \ \ cb}\, ,
\ee
we have
\be                                                     \label{c9aa}
K^{(1)}_{ab}=\eta^{(1)}_{ab},\ \ \ \ 
K^{(2)}_{ab}=-s^2\eta^{(2)}_{ab}\, .
\ee
As a result, the metric for the SU(2) part of the internal space is 
(proportional to)
the corresponding Cartan metric, while the one for the SU(1,1) factor is the 
{\em negative} Cartan metric.

Let us now describe  the structure of the vielbein in D=10. We have
\be                                                     \label{c14}\
\hat{{g}}_{MN}d\hat{x}^M d\hat{x}^N=
\hat{\eta}_{AB}
\hat{\Theta}^{A}\hat{\Theta}^{B}=
\eta_{\alpha\beta}\hat{\Theta}^{\alpha}\hat{\Theta}^{\beta}
+\sum_{(\sigma)=1,2}
\eta^{(\sigma)}_{ab}\hat{\Theta}^{(\sigma) a}\hat{\Theta}^{(\sigma) b},
\ee
where 
\be                                                     \label{c15}
\hat{\Theta}^{\alpha}={\rm e}^{-3\phi/4}\,\theta^{\alpha}_{\ \mu}dx^\mu,
\ \ \ \
\hat{\Theta}^{(\sigma)a}=
\left.\left.
\frac{\sqrt{2}}{g_{(\sigma)}}\,
{\rm e}^{\phi/4}
\right( A^{(\sigma)a}_{\mu}dx^\mu-\tilde{\theta}^{(\sigma)a}\right) .
\ee
The dual basis $\hat{E}_B$ is specified by 
\be                                                     \label{c16}
\hat{{g}}^{MN}
\frac{\partial}{\partial\hat{x}^M}
\frac{\partial}{\partial\hat{x}^N}=\hat{\eta}^{AB}
\hat{E}_{A}\hat{E}_{B}=
\eta^{\alpha\beta}\hat{E}_{\alpha}\hat{E}_{\beta}
+\sum_{(\sigma)=1,2}
\eta^{(\sigma)ab}\hat{E}^{(\sigma)}_{a}\hat{E}^{(\sigma)}_{b},
\ee
where
\bea                                            \nonumber        
\hat{E}_\alpha={\rm e}^{3\phi/4}
\left(e_\alpha+\sum_{(\sigma)=1,2}e_\alpha^{\ \mu}
A^{(\sigma)a}_{\mu}\tilde{e}^{(\sigma)}_a\right),\ \ \ \
\hat{E}^{(\sigma)}_a\equiv\hat{E}_{\rm a}
=-\frac{g_{(\sigma)}}{\sqrt{2}}\,
{\rm e}^{-\phi/4}\,\tilde{e}^{(\sigma)}_{a} . \\
                                      \label{c17}
\eea
Here $e_\alpha\equiv e_\alpha^{\ \mu}\,\partial/\partial x^\mu$
are the basis 4-vectors dual to the $\theta^\alpha$'s, 
and 
$\tilde{e}^{(\sigma)}_{a}\equiv
\tilde{e}^{(\sigma)i}_{a}\partial/\partial z^{(\sigma)i}$
are the invariant vectors on the group spaces ${\cal G}^{(\sigma)}$.

We shall also need explicit expressions for $\hat{R}_{MN}$. 
These can be obtained in the standard way from the vielbein
connection $\hat{\omega}_{AB,C}$, which is computed 
in Eq. (\ref{d27}) below. This yields components 
$\hat{R}_{AB}=
(\hat{R}_{\alpha\beta},\hat{R}_{\alpha\rm a},\hat{R}_{\rm ab})$
in the basis (\ref{c17}):
\bea
&&{\rm e}^{-3\phi/2}\,
\hat{R}_{\alpha\beta}\,\theta^\alpha_{\ \mu}\theta^\beta_{\ \nu}=
R_{\mu\nu}-\frac32\,\partial_\mu\phi\,\partial_\nu\phi
+\frac34\,g_{\mu\nu}\nabla_\rho\nabla^\rho\phi -\frac12\,{\rm e}^{2\phi}
\gamma_{\rm ab}\,F^{\rm a}_{\ \mu\rho}
F^{\rm b\  \rho}_{\ \,\nu}\, ,\nonumber                        \\ 
&&\frac{\sqrt{2}}{g_{\rm (a)}}\,{\rm e}^{-\phi/2}\,
\hat{R}_{{\rm a}\alpha}\,\theta^\alpha_{\ \mu}=-\left.\left.\frac12\,
\gamma_{\rm ab} \right(\nabla_\rho({\rm
e}^{2\phi}F^{{\rm b}\rho}_{\ \ \, \mu})
+ {\rm e}^{2\phi}\,f^{\rm b}_{\rm \ \
cd}A^{\rm c}_\rho  F^{{\rm d}\rho}_{\ \ \, \mu}\right)\, ,  \label{c9} \\
&&\frac{8}{g_{\rm (a)}g_{\rm (b)}}\,{\rm e}^{\phi/2}\,
\hat{R}_{\rm ab}=-{\rm e}^{2\phi}\,
\gamma_{\rm ab}\,\nabla_\rho\nabla^\rho
\phi\,+{\rm e}^{4\phi}\,
\gamma_{\rm ac}\gamma_{\rm bd}\,
F^{\rm c}_{\ \mu\nu}F^{\rm d\mu\nu}+
f^{\rm d}_{\rm \ \, ac}f^{\rm c}_{\rm \ \, db}\, ,   \nonumber
\eea 
where $\nabla_\rho$ and $R_{\mu\nu}$ are the covariant derivative and 
Ricci tensor for $g_{\mu\nu}$, and $g_{\rm (a)}\equiv
g_{(\sigma)a}=g_{(\sigma)}$.

So far we have expressed the ten-dimensional
$\hat{g}_{MN}$ and $\hat{\phi}$  in terms of the
four-metric $g_{\mu\nu}$, the dilaton $\phi$, and the 
gauge fields $A^{(\sigma)a}_{\mu}$.
It remains to specify the ten-dimensional  antisymmetric
tensor $\hat{H}_{MNP}$. 
We choose
its non-vanishing components  in the basis (\ref{c17}) to be 
\cite{Chamseddine98}
\bea
\hat{H}^{(\sigma)}_{abc}&=&\frac{g_{(\sigma)}}{2\sqrt{2}}\,
{\rm e}^{-3\phi/4}\,\varepsilon_{abc}\, ,              \nonumber     \\
\hat{H}^{(\sigma)}_{a\alpha\beta}&=&-\frac{1}{\sqrt{2}g_{(\sigma)}}\,
{\rm e}^{5\phi/4}\,
\eta^{(\sigma)}_{ab}F^{(\sigma)b}_{\ \alpha\beta}\, ,\nonumber    \\
\hat{H}_{\alpha\beta\gamma}&=&{\rm e}^{-7\phi/4}\,
\varepsilon_{\alpha\beta\gamma}^{\ \ \ \ \,\delta}\,
e_\delta^{\ \mu}\partial_\mu{\bf a}\, .   \label{c18}
\eea
Here the axion ${\bf a}$ depends only on $x^\mu$, 
we choose $\varepsilon^{0123}=1$, and 
$F^{(\sigma)b}_{\ \ \ \alpha\beta}$ are components of the gauge
field strength with respect to the tetrad $e_\alpha$.

\subsection{Four-dimensional theory}

We now have  the complete ansatz for 
$\hat{g}_{MN}$,  $\hat{H}_{MNP}$ and $\hat{\phi}$,  
and this
we insert into the supergravity equations 
(\ref{c1})--(\ref{c3}). Let us consider first Eq. (\ref{c1}) for the 
dilaton $\hat{\phi}$. Using the above definitions, it is not 
difficult to see that this equation assumes the following 
four-dimensional form: 
\be                                                     \label{c19}
\nabla_\rho\nabla^\rho\phi=\frac12\,
{\rm e}^{2\phi}\sum_{(\sigma)=1,2}\frac{1}{g_{(\sigma)}^2}\,
\eta^{(\sigma)}_{ab}
F^{(\sigma)a}_{\ \,\mu\nu}F^{(\sigma)b \mu\nu}+2s^2{\rm e}^{-4\phi}\,
\partial_\rho{\bf a}\,\partial^\rho{\bf a} -2{\rm U}(\phi),
\ee
with the dilaton potential 
\be                                                     \label{c20}
{\rm U}(\phi)=-\left.\left.\frac18\,
\right(g_{(1)}^2-s^2g_{(2)}^2\right){\rm e}^{-2\phi}\, .
\ee
The next step is to check Eq. (\ref{c2}) for $\hat{H}_{MNP}$. 
This is in fact a system of 49 equations labeled by pairs of indices $(M,N)$,
and it can be split into three groups labeled by $(\mu,\nu)$, (a,b), 
and $(\mu$,a), respectively. A direct computation then reveals that 
the $(\mu,\nu)$ and (a,b) equations are identically fulfilled, while 
the $(\mu$,a) ones reduce to 
\be                                                     \label{c21}
\nabla_\rho({\rm e}^{2\phi}F^{(\sigma)a\rho\mu})
+{\rm e}^{2\phi}f_{\ \ \ \ bc}^{(\sigma) a}A^{(\sigma) b}_\rho
F^{(\sigma)c\rho\mu}=
2\ast\!F^{(\sigma)a\mu\rho}\partial_\rho{\bf a}\, .
\ee
These are the four-dimensional Yang-Mills equations; the 
dual tensor is defined as $\ast\! F^{(\sigma)a}_{\mu\nu}=
\frac12\sqrt{|g|}\varepsilon_{\mu\nu\lambda\rho}F^{(\sigma)a\lambda\rho}$.

Let us now turn to the Einstein equations in (\ref{c3}). Splitting 
these into the $(\mu,\nu)$, (a,b) and $(\mu$,a) groups and 
using the expressions for $\hat{R}_{MN}$ in (\ref{c9}) together with 
all the above definitions, we find that the (a,b) equations are identically
fulfilled, while the $(\mu$,a) ones again reduce  to the Yang-Mills equations
(\ref{c21}). The $(\mu,\nu)$ group gives the 
four-dimensional Einstein equations:
\be                                           \label{c22}
R_{\mu\nu}=2\,\partial_\mu\phi\,\partial_\nu\phi
-2s^2{\rm e}^{-4\phi}\partial_\mu{\bf a}\,\partial_\nu{\bf a}
+2\,T_{\mu\nu}+2{\rm U}(\phi)\,g_{\mu\nu}\, , \\
\ee
with the Yang-Mills energy-momentum tensor 
\be                                     \label{c23}
T_{\mu\nu}={\rm e}^{2\phi}\sum_{(\sigma)=1,2}
\frac{1}{g_{(\sigma)}^2}\,
\eta^{(\sigma)}_{ab}
\left(F^{(\sigma)a}_{\ \ \ \ \mu\rho}
F^{(\sigma)b\ \rho}_{\ \ \ \ \, \nu}
-\frac14\,g_{\mu\nu}
F^{(\sigma)a}_{\ \ \ \ \lambda\rho}
F^{(\sigma)b\lambda\rho}\right) .  
\ee
At this point our procedure successfully terminates, since {\it all}
ten-dimensional equations are now fulfilled,
provided that the four-dimensional
dilaton, Yang-Mills, and Einstein equations in (\ref{c19})--(\ref{c23}) hold. 
We note, however,  that so far we have not obtained
the equation for the axion ${\bf a}$. 
This arises as the consistency condition for the Einstein equations
(\ref{c22}). Specifically, the Bianchi identities for (\ref{c22})
are fulfilled
by virtue of the dilaton and Yang-Mills equations together with 
\be            
\nabla_\rho\nabla^\rho ({\rm e}^{-4\phi}{\bf a})=\frac{s^2}{2}\,
\sum_{(\sigma)=1,2}\frac{1}{g_{(\sigma)}^2}\,\eta^{(\sigma)}_{ab}
\ast\! F^{(\sigma)a}_{\ \,\mu\nu}F^{(\sigma)b\mu\nu}\, . \label{c24}
\ee
This completes the system of four-dimensional equations. 
It is not difficult to see that all equations in (\ref{c19})--(\ref{c24}) 
can be obtained by varying the four-dimensional action
\be                                                        \label{c25}
S_4=\int{\cal L}_4\sqrt{|g|}\, d^4x\, ,
\ee
where
\bea
{\cal L}_4&=&\frac{R}{4}\,-\frac12\,\partial_\mu\phi \,\partial^\mu\phi
+\frac{s^2}{2}\,{\rm e}^{-4\phi}\, 
\partial_\mu{\bf a}\,\partial^\mu{\bf a}  
-\frac14\,
{\rm e}^{2\phi}\sum_{(\sigma)=1,2}\frac{1}{g_{(\sigma)}^2}\,
\eta^{(\sigma)}_{ab}
F^{(\sigma)a}_{\ \,\mu\nu}F^{(\sigma)b \mu\nu}  \nonumber  \\
&-&\frac{1}{2}\,{\bf a}
\sum_{(\sigma)=1,2}\frac{1}{g_{(\sigma)}^2}\,\eta^{(\sigma)}_{ab}
\ast\! F^{(\sigma)a}_{\ \,\mu\nu}F^{(\sigma)b\mu\nu}
+\left.\left.\frac18\,                                  
\right(g_{(1)}^2-s^2g_{(2)}^2\right){\rm e}^{-2\phi}  \, . \label{c26}   
\eea
We have therefore obtained a four-dimensional theory via the consistent
dimensional reduction of ten-dimensional supergravity. 
In fact, we have obtained two different theories 
distinguished by the values of the parameter
$s=1,i$.
Let us recall that  the signature
of the spacetime metric
is $(s^2,+1,+1,+1)$,
the internal metrics are 
$\eta^{(2)}_{ab}={\rm diag}(+1,+1,-s^2)$
and $\eta^{(1)}_{ab}=\delta_{ab}$, and  
the field tensors $F^{(\sigma)a}_{\ \,\mu\nu}$ are specified by
Eqs. (\ref{c11a}), (\ref{c13}). 

The model (\ref{c26})
describes interacting gravitational, axion, dilaton,
and two non-Abelian gauge fields with  gauge group ${\cal G}$ 
and two independent gauge coupling
constants $g_{(1)}$ and $g_{(2)}$. For $s=i$ the gauge group is
SU(2)$\times$SU(2), and the theory coincides 
with the bosonic sector of the gauged supergravity of Freedman and
Schwarz \cite{Freedman78}. 
We have therefore reproduced the result of
Ref.\cite{Chamseddine98} that the Freedman-Schwarz model 
can be obtained via dimensional reduction of D=10, N=1 supergravity
on $S^3\times S^3$. The inverse radii of the spheres 
determine the gauge coupling constants $g_{(\sigma)}$. 

The principal new result that we obtain here is the model determined
by (\ref{c26}) for $s=1$. This has been hitherto unknown. 
It is somewhat similar to the Freedman-Schwarz model, apart from the 
fact that it lives in Euclidean and not Lorentzian space,  also 
the gauge group is now   SU(2)$\times$SU(1,1), and the dilaton potential 
is proportional to the difference and not to 
the sum of the coupling constants. 
The latter is due to the opposite signs of the scalar curvatures 
of the group manifolds used for the dimensional reduction:
SU(2) with the positive metric and SU(1,1) with the negative metric. 

We call the new theory Euclidean Freedman-Schwarz 
(EFS) model.  Some of its features are as follows. We notice
that the kinetic terms for the dilaton and axion in (\ref{c26}) 
have opposite signs, which is typical for  Euclidean theories. 
We notice also that the non-compact Lie-algebra components
of  $F^{(2)a}_{\ \,\mu\nu}$  $(a=1,2)$ give positive
contributions to the energy density, while the compact one 
$(a=3)$ makes
a negative contribution. This is because
the metric for bilinear combinations of the gauge field strength 
is not the Cartan metric for SU(1,1) but its negative, $\eta^{(2)}_{ab}$.  
So far the new supergravity is not yet complete, since we have described 
only its bosonic sector. We shall now pass to considering the 
fermions, in which we shall restrict ourselves to deriving
the linearized fermion supersymmetry transformations.

\section{Fermions}
\setcounter{equation}{0}

The fermion fields of D=10, N=1 supergravity,
the gravitino $\hat{\psi}_M$
and the gaugino $\hat{\chi}$, can be consistently set to zero,
which leads to the action (\ref{c0}). However,
their supersymmetry variations do not
necessarily vanish and are given by
\bea
\delta \hat{\psi }_{M}&=&\hat{\cal D}_M\,\hat{\epsilon}
-\frac{1}{48}\, e^{-\hat{\phi}}\,           
\left( \hat{\Gamma}_{\ \ \ \  M}^{SPQ}+9\,\delta _{M}^{S}\,
\hat{\Gamma}^{PQ}\right) \,\hat{H}_{SPQ}\,\hat{\epsilon}\, , \label{c4}  \\
\delta \hat{\chi }&=&-\frac{1}{\sqrt{2}}\, 
(\hat{\Gamma}^{M}\partial _{M}\hat{\phi})\,\hat{\epsilon}
-\frac{1}{12\sqrt{2}}\,e^{-\hat{\phi}}%
 \,\hat{\Gamma}^{SPQ}\, \hat{H}_{SPQ}\,\hat{\epsilon}\, .  \label{c5}
\eea
Here $\hat{\epsilon}$ is the Majorana-Weyl spinor parameter  
of supersymmetry transformations, and its covariant derivative is
\be                                                     \label{c5a}
\hat{\cal D}_{M}\,\hat{\epsilon}=\left(\partial_M+
\frac14\,\hat{\omega}_{AB,M}\,\hat{\Gamma}^{AB}\right)\hat{\epsilon}\, ,
\ee
with $\partial_M\equiv\partial/\partial\hat{x}^M$ and
$\hat{\omega}_{AB,M}$  being the spin connection
for the vielbein $\hat{E}_{A}$.
The D=10 gamma matrices span the Clifford algebra
$\hat{\Gamma}^{A}\hat{\Gamma}^{B}
+\hat{\Gamma}^{A}\hat{\Gamma}^{B}
=2\,\hat{{\bf \eta}}^{AB}$; one has 
$\hat{\Gamma}^{M}\equiv\hat{\Gamma}^{A}\hat{E}^{\ M}_{A}$ and 
$\hat{\Gamma}^{M\ldots N}
\equiv\hat{\Gamma}^{[M}\ldots \hat{\Gamma}^{N]}$.

We shall now proceed as in the bosonic case to express the 
ten-dimensional quantities in terms  of the four-dimensional ones. 
Our aim is to consistently
derive the four-dimen\-sio\-nal supersymmetry transformations
from the ten-dimensional rules (\ref{c4}),(\ref{c5}).
The vanishing
of the four-dimensional SUSY variations will then  imply that 
the ten-dimensional variations 
$\delta \hat{\psi }_{M}$ and $\delta \hat{\chi}$  vanish.  

\subsection{The D=10 gamma matrices}
We parameterize
the 32$\times$32 gamma matrices 
$\hat{\Gamma}^A\equiv
(\hat{\Gamma}^\alpha, \hat{\Gamma}^{(1)a}, \hat{\Gamma}^{(2)a})$ as
\bea
\hat{\Gamma}^\alpha&=&
{\bf \1}_2\otimes\gamma^\alpha\otimes\1_4\, ,\nonumber \\
\hat{\Gamma}^{(1)a}&=&
i\,\mbox{\boldmath $\tau$}^1\otimes\gamma_5
\otimes\alpha^{(1)a}\, ,\nonumber \\
\hat{\Gamma}^{(2)a}&=&
\frac1s\,\mbox{\boldmath $\tau$}^3
\otimes\gamma_5\otimes\alpha^{(2)a}\, .\label{d1}
\eea
Here as usual $s=1,i$; 
$\mbox{\boldmath $\tau$}^a$ are the Pauli matrices, $\gamma^\alpha$ are the   
D=4 gamma matrices, and the $4\times 4$ matrices $\alpha^{(\sigma)a}$ 
generate the Lie algebra of the  group ${\cal G}$. 
One has 
\be                                                     \label{d2}
\gamma^\alpha\gamma^\beta
+\gamma^\beta\gamma^\alpha=2\,\eta^{\alpha\beta}
\equiv 2\,{\rm diag}(s^2,+1,+1,+1)\, .
\ee
Since $\varepsilon^{0123}=1$ and $\varepsilon_{0123}=s^2$,
and also $\sqrt{-\eta}=is$, we have
\be                                                     \label{d3}
\gamma_5=\frac{i}{4!}\,\sqrt{-\eta}\,\varepsilon_{\alpha\beta\gamma\delta}\,
\gamma^\alpha\gamma^\beta\gamma^\gamma\gamma^\delta=
-\frac1s\,\gamma^0\gamma^1\gamma^2\gamma^3\, ,
\ee
such that $\gamma_{5}^2=1$ and $\{\gamma_5,\gamma^\alpha\}=0$. 
The matrices $\alpha^{(\sigma)a}$ in (\ref{d1})
are specified by the relations
\bea                                                    \label{d5}
&&\alpha^{(\sigma)a}\alpha^{(\sigma)b}=
-\varepsilon^{abc}\,\eta^{(\sigma)}_{cd}\alpha^{(\sigma)d}
- K^{(\sigma)ab}\,  ,  \nonumber  \\
&&\left(\alpha^{(\sigma)a}\right)^\dagger
=- K^{(\sigma)ab}\,\delta_{bc}\,
\alpha^{(\sigma)c}\, ,\ \  \nonumber \\
&&[\alpha^{(1)a},\alpha^{(2)b}]=0\, ,
\eea
where $\varepsilon^{abc}=\varepsilon_{abc}$ and
the Cartan metric $K^{(\sigma)}_{ab}$   
is defined in Eq. (\ref{c9a}). For
$\T^{(\sigma)}_a=-\frac12\,\delta_{ab}\,\alpha^{(\sigma)b}$  we have
\be                                                     \label{d6}
[\T^{(\sigma)}_a,\T^{(\sigma)}_b]=\varepsilon_{abc}\,\eta^{(\sigma)cd} \, 
\T^{(\sigma)}_d\, ,
\ee
which are the commutation relations for the Lie algebra
of  ${\cal G}$. 

It is not difficult to see that Eqs. (\ref{d1})--(\ref{d5}) imply the correct 
Clifford algebra relations for the $\hat{\Gamma}^A$'s. 
Although the actual choice of $\gamma^\alpha$ and 
$\alpha^{(\sigma)a}$ is not important,
it is sometimes convenient to have an explicit representation.
One can choose
\be                                                     \label{d4}
\gamma^0=s\,\sigma^1\otimes\1_2, \ \
\gamma^{\ua}=\sigma^2\otimes\sigma^{\ua},\ \
\gamma_5=\sigma^3\otimes\1_2\, ,
\ee
and also 
\be                                                     \label{d7}
\alpha^{(1)a}=i\,\tau^a\otimes\1_2\, ;\ \ \
\alpha^{(2)b}=s\,\1_2\otimes\tau^b\,\ \ (b=1,2),\ \ 
\alpha^{(2)3}=i\1_2\otimes\tau^3\, ,
\ee
with $\sigma^a$ and $\tau^a$ being  Pauli matrices.

\subsection{The Majorana-Weyl condition}
Having expressed the $\hat{\Gamma}^A$'s in terms of
four-dimensional quantities, we need a similar reduction also
for the spinors. It is important that the
spinors in Eqs. (\ref{c4}), (\ref{c5}) are Majorana-Weyl with
$\delta\hat{\psi}_S$ and $\hat{\epsilon}$ 
being right handed, while $\delta\hat{\chi}$
is left-handed \cite{Chamseddine81}.  
Upon  dimensional reduction 
the D=10 Majorana-Weyl condition will 
reduce to Majorana-type constraints for the 
D=4 spinors,
which we shall discuss in some detail, 
especially in the Euclidean case.

Let us first consider the Weyl condition. As is well-known, this
can be imposed in any even-dimensional space. In the particular case
of D=10 one defines the chirality matrix as
\be                                                     \label{d8}
\hat{\Gamma}_{11}=-\hat{\Gamma^0}\hat{\Gamma^1}\ldots
\hat{\Gamma^9}=
\mbox{\boldmath$\tau$}^2\otimes\gamma_5\otimes\1_4\, ,
\ee
the Weyl spinors $\hat{\psi}_\pm$ being solutions of 
\be                                                     \label{d9}
\hat{\Gamma}_{11}\hat{\psi}_\pm=\pm\hat{\psi}_\pm\, .
\ee
Let us now consider the Majorana condition. This is defined by 
the matrix $\hat{B}$ subject to
\be                                                     \label{d11}
\hat{B}\,\hat{\Gamma}^A\hat{B}^{-1}=\left(\hat{\Gamma}^A\right)^\ast\, ,
\ee
where the asterisk denotes complex conjugation. Such a matrix 
always exists, since $\hat{\Gamma}^A\to(\hat{\Gamma}^A)^\ast$ 
is a symmetry
of the Clifford algebra. This implies that 
for any spinor $\hat{\psi}$ in any dimension D 
and for an arbitrary spacetime signature  
one can define the conjugated spinor 
$\hat{\psi}_{\rm M}\equiv\hat{B}^{-1}\hat{\psi}^\ast$, which transforms 
as $\hat{\psi}$. 
What is not always possible is to choose  $\hat{B}$ such that 
\be                                                     \label{d13}     
\hat{B}\hat{B}^\ast=1\, .
\ee
As a result, the Majorana condition, $\hat{\psi}_{\rm M}=\hat{\psi}$, 
or explicitly
\be                                                     \label{d12}
\hat{\psi}^\ast=\hat{B}\hat{\psi}\, ,
\ee  
is not always consistent, since it requires that (\ref{d13}) must hold. 
For example, for D=4 the solution to (\ref{d11}), (\ref{d13}) 
exists in Minkowski space
but not in Euclidean space. In those cases where
the Majorana condition can be imposed, it is not always compatible
with the Weyl condition, since chirality may change under the Majorana
conjugation. 

As is well-known, in ten-dimensional space with the signature (1,9) the
Majorana condition can be imposed and is compatible with the Weyl
condition. Let us see this explicitly.
In the representation (\ref{d1}) one has
\be                                                     \label{d14}
\hat{B}=\mbox{\boldmath $\Omega$}\otimes A\otimes B\, ,
\ee
where
\be                                                     \label{d15}
\mbox{\boldmath $\Omega$}=
\left(\begin{array}{cc}
1 & 0\\
0 & -s^2
\end{array}\right), \ \ \ 
\ee
and the $4\times 4$ matrices $A$ and $B$ are such that
\be                                                     \label{d16}
A\gamma^\alpha A^{-1}=\left(\gamma^{\alpha}\right)^\ast,\ \ \
B\alpha^{(\sigma)a} B^{-1}=\left(\alpha^{(\sigma)a}\right)^\ast\, ,
\ee
and also
\be                                                     \label{d16a}
AA^\ast\otimes BB^\ast=\1\, .
\ee
It follows that $A\gamma_5 A^{-1}=s^2\,\gamma_5^\ast$.
The explicit form of $A$ and $B$ depends on the representation 
of $\gamma^\alpha$ and $\alpha^{(\sigma)a}$.
In the representation (\ref{d4}),
(\ref{d7}) one has
\be                                                     \label{d20}
A=\1_2\otimes\sigma^2,\ \ \ \ \ B=\tau^2\otimes\tau^1\, 
\ee
for $s=1$, while for $s=i$ one finds
\be
A=\sigma^2\otimes\sigma^2,\ \  \ \ \ B=\tau^2\otimes\tau^2\, .
\ee
Using these definitions 
one can see that the Weyl condition in (\ref{d8}),(\ref{d9}) and
the Majorana condition in (\ref{d12})
can be solved simultaneously as
\be                                                     \label{d10}
\hat{\psi}_\pm=
\left(\begin{array}{c}
\psi \\
\pm i\gamma_5\psi
\end{array}\right)\, ,
\ee
with
\be                                                     \label{d17}
\psi^\ast=A\otimes B\,\psi\, .
\ee
Here $\hat{\psi}$ is written in the form of a two-component spinor
which is acted upon by the $2\times 2$ bold-faced matrices like
${\bf \1}_2$, $\mbox{\boldmath$\tau$}^a$ and  
$\mbox{\boldmath$\Omega$}$. The spinor $\psi$ 
has 16 components, which are acted upon by the $4\times 4$
gamma-matrices $\gamma^\alpha$ and group-generators  
$\T^{(\sigma)}_a$. One can write 
$\psi\equiv\psi^{\rm I}_\kappa$, where I$=1,\ldots 4$ is 
the group index and 
$\kappa=1,\dots 4$ is the spinor index.  In view of  (\ref{d17})
only 8 components of $\psi$ are independent. We notice that 
the condition  (\ref{d17}) is invariant under
4-rotations generated by
$\frac14[\gamma^\alpha,\gamma^\beta]$ and gauge
transformations generated by $\T^{(\sigma)}_a$. As a result, $\psi$
can be viewed as a ${\cal G}$-group multiplet 
of Majorana spinors in D=4 with the Majorana condition 
given by 
(\ref{d17}). Eqs. (\ref{d10}), (\ref{d17})  therefore provide
the sought expression for the ten-dimensional spinors
in terms of the four-dimensional ones. 

The Majorana condition (\ref{d17}) is obtained
for both values of the parameter $s$. 
One might think that for $s=1$, when the space is Euclidean,  
this contradicts the well-known fact  
that there are no Majorana spinors
in four-dimensional Euclidean space. 
However, there is no contradiction, since the spinors
have  internal degrees of freedom. 
Specifically, 
the normalization (\ref{d16a}), which is the analog of (\ref{d13}), 
holds because there are {\it two} factors
on the left hand side of (\ref{d16a}), which satisfy
\be                                                     \label{d21}
AA^\ast=-s^2,\ \ \ \ \ BB^\ast=-s^2 \, .
\ee
For $s=1$ each of the two terms here has the wrong sign, but two wrongs
make a right -- their product in  (\ref{d16a})  has the correct sign.
Without the matrix
$B$ one would be left with just one wrong sign, and this implies that
singlet fermions cannot be
Majorana. To recapitulate, the 
Euclidean Majorana condition (\ref{d17}) is consistent due to the 
group degrees of freedom. We also note  that, since 
$A\gamma_5 A^{-1}=+(\gamma_5)^\ast$ for $s=1$, 
the Majorana spinors can be at the same time Weyl.

For $s=i$ each of the two factors in
(\ref{d16a}) has the correct sign on its own. 
In particular, for $s=i$ one can choose all the $\alpha^{(\sigma)a}$'s 
to be real \cite{Freedman78}, in which case $B=1$. 
As a result, the Majorana condition 
can be imposed both for group singlets and multiplets. 
However, since 
$A\gamma_5 A^{-1}=-(\gamma_5)^\ast$, spinors cannot at the same time
be  Majorana and Weyl.

\subsection{Four-dimensional SUSY variations} 
We  now have all  necessary tools in order  
to reduce the D=10 SUSY transformations in 
(\ref{c4}), (\ref{c5}) to four dimensions.
Let us first consider  Eq. (\ref{c5}) for $\delta\hat{\chi}$. 
The spinors $\delta\hat{\chi}$ and $\hat{\epsilon}$
have left and right chiralities, respectively, and we therefore choose
according to (\ref{d10})
\be                                     \label{d22}
\delta\hat{\chi}=
-\frac12\,{\rm e}^{5\phi/8}
\left(\begin{array}{c}
\delta\chi\\
-i\gamma_5\,\delta\chi
\end{array}\right), \ \ \ 
\hat{\epsilon}=
{\rm e}^{-\phi/8}
\left(\begin{array}{c}
\epsilon \\
i\gamma_5\,\epsilon
\end{array}\right). \ \ \ 
\ee
Inserting this into (\ref{c5}), using
the $\Gamma^A$'s 
from (\ref{d1}) and $\hat{H}_{ABC}$ from (\ref{c18}), 
and also utilizing the identity
\be                                             \label{d23}
\gamma_\alpha\gamma_5=
\frac{i}{6}\,\sqrt{-\eta}\,\varepsilon_{\alpha\beta\gamma\delta}\,
\gamma^\beta\gamma^\gamma\gamma^\delta\, ,
\ee
Eq. (\ref{c5}) reduces to the following relation between $\delta\chi$ and 
$\epsilon$:
\bea                                            \label{d24}
\delta\chi&=&\left(\frac{1}{\sqrt{2}}\,\gamma^\mu\partial_\mu\phi-
\frac{1}{\sqrt{2}s}\,{\rm e}^{-2\phi}
\gamma_5\gamma^\mu\partial_\mu{\bf a}\right)
\epsilon \nonumber \\
&+&\frac{1}{2s}\,{\rm e}^\phi
\left(s{\cal F}^{(1)}-\gamma_5{\cal F}^{(2)}\right)
\epsilon
+\frac{1}{4s}\,{\rm e}^{-\phi} \left(s\,g_{(1)}-
g_{(2)}\gamma_5\right)\epsilon\, ,
\eea
with
\be                                                     \label{d25}
{\cal F}^{(\sigma)}=\frac{1}{2g_{(\sigma)}}\,
\eta^{(\sigma)}_{ab}\,\gamma^\alpha\gamma^\beta\,
F_{\alpha\beta}^{(\sigma)a}\alpha^{(\sigma)b}\, .
\ee
We note that this relation is compatible with the 
Majorana condition (\ref{d17}) for the spinors
$\delta\chi$ and $\epsilon$. 
Specifically, taking the Majorana conjugate of 
(\ref{d24}) and using (\ref{d16}), the whole expression reproduces itself. 

Consider now  the equation for $\delta\hat{\psi}_M$ in (\ref{c4}). 
The procedure in this case is somewhat more involved. 
The first step is to compute the spinor covariant
derivatives in (\ref{c5a}), and for this we need the spin-connection
$\hat{\omega}_{AB,M}$. This can be obtained as
\be                                                     \label{d26}
\hat{\omega}_{AB,C}=\frac12\,(C_{B,AC}+C_{C,AB}-C_{A,BC})\, ,
\ee
where $C_{A,BC}=\hat{\eta}_{AD}C^D_{\ BC}$ are determined by the 
commutation relations for the basis vectors of the vielbein (\ref{c17}),
\be
[\hat{E}_A,\hat{E}_B]=C^C_{\ AB}\,\hat{E}_C\, .
\ee
The result is
\bea
\hat{\omega}_{\alpha\beta,\gamma}&=&
{\rm e}^{3\phi/4}\left(\omega_{\alpha\beta,\gamma}
+\frac34\,
(\eta_{\beta\gamma}\,e_{\alpha}^{\ \mu}-
\eta_{\alpha\gamma}\,e_{\beta}^{\ \mu})\,\partial_\mu\phi\right)\, ,\ \  
\nonumber \\
\hat{\omega}_{\alpha\beta,a}^{(\sigma)}&=&
\hat{\omega}_{\alpha a,\beta}^{(\sigma)}=
-\frac{1}{\sqrt{2}g_{(\sigma)}}\,{\rm e}^{7\phi/4}\,
\eta^{(\sigma)}_{ab}\,F^{(\sigma)b}_{\alpha\beta}\, ,  \nonumber \\
\hat{\omega}_{\alpha a,b}^{(\sigma)}&=&
-\frac{1}{4}\,{\rm e}^{3\phi/4}\,\eta_{ab}\,e_\alpha^{\ \mu}\partial_\mu\phi\, 
,\ \ \
\hat{\omega}_{ab,\alpha}^{(\sigma)}=
-{\rm e}^{3\phi/4}\,\varepsilon_{abc}\,A^{(\sigma)c}_\alpha
\, ,\ \ \ \nonumber \\
\hat{\omega}_{ab,c}^{(\sigma)}&=&
\frac{g_{(\sigma)}}{2\sqrt{2}}\,
{\rm e}^{-\phi/4}\,\varepsilon_{abc}\, ,\ \ \    \label{d27}
\eea
where $\omega_{\alpha\beta,\gamma}$ is the spin-connection for the 
tetrad $e_\alpha$.

Using these expressions together with (\ref{d1}) and (\ref{c18})
we first consider 
that part of the SUSY variation $\delta\hat{\psi}_M$
for which the index 
runs over the internal coordinates, $M$=m.  Utilizing the identity
\be                                                     \label{d28}
\hat{\Gamma}^{ABC}_{\ \ \ \ \ D}\,\hat{H}_{ABC}=
(-\hat{\Gamma}_D\hat{\Gamma}^{ABC}
+3\delta^A_D\,\hat{\Gamma}^{BC})\,
\hat{H}_{ABC}\, ,
\ee
a straightforward computation gives the relation for the D=10 spinors,
\be                                                     \label{d29}
\delta\hat{\psi}_{\rm a}+\frac{1}{2\sqrt{2}}\,
\hat{\Gamma}_{\rm a}\,\delta\hat{\chi}
=-\frac{g_{\rm (a)}}{\sqrt{2}}\,{\rm e}^{-\phi/4}\,
\tilde{e}_{\rm a}^{\rm m}\,
\frac{\partial}{\partial z^{\rm m}}\,\hat{\epsilon}\, ,
\ee
where $g_{\rm (a)}$ is defined after Eqs. (\ref{c9})
(one has $\delta\hat{\psi}_{\rm m}=
\hat{\Theta}^{A}_{\ \rm m}\delta\hat{\psi}_{A}=
\hat{\Theta}^{\rm a}_{\ \rm m}\delta\hat{\psi}_{\rm a}$). 
Notice the following important fact: the expression on the right
contains the partial and not the covariant
derivative.
Specifically, when simplifying the expression on the left 
the  spin connection 
arises twice, and a careful examination reveals that the
two terms cancel each other. 
As a result, the expression in (\ref{d29}) appears to be
not covariant under 
local rotations of the internal basis $\tilde{e}_{\rm a}$.
This, however, is simply a consequence of the fact  
that the whole theory under consideration
does not allow for such  rotations. Indeed, the 
crucial assumption is that the basis $\tilde{e}_{\rm a}$ consists of 
{\it invariant} vector fields, for which 
only  global rotations are allowed. 
Under these, obviously, Eq. (\ref{d29}) is covariant.

Consider now the four-dimensional part of the gravitino,
$\delta\hat{\psi}_\mu$.
Since $\delta\hat{\psi}_M$ has positive chirality, 
we consider the linear combination 
\cite{Chamseddine81,Chamseddine98}
\be                                                     \label{d30}
\delta\hat{\psi}_\mu-
\frac{3}{2\sqrt{2}}\,\gamma_\mu\,\delta\hat{\chi}\equiv 
{\rm e}^{5\phi/8}\,
\left(\begin{array}{c}
\delta\psi_\mu \\
i\gamma_5\,\delta\psi_\mu
\end{array}\right).
\ee 
Taking into account all the definitions above and making use of the 
identity
\be                                                             \label{d31}
\gamma_\alpha\gamma_\beta\gamma_5=
-\frac{i}{2}\,\sqrt{-\eta}\,\varepsilon_{\alpha\beta\gamma\delta}\,
\gamma^\gamma\gamma^\delta+\eta_{\alpha\beta}\gamma_5\, ,
\ee
the remaining part of the SUSY variation in (\ref{c4}) 
reduces to the following relation:
\bea                                                    \label{d32}
&&\delta\psi_\mu=\left(\partial_\mu
+\frac14\,\omega_{\alpha\beta,\mu}\gamma^\alpha\gamma^\beta
-\frac12\,\sum_{(\sigma)=1,2}K^{(\sigma)}_{ab}
\alpha^{(\sigma)a}A_{\mu}^{(\sigma)b}+
\frac{1}{2s}\,{\rm e}^{-2\phi}\gamma_5\,\partial_\mu{\bf a}\right)
\epsilon \nonumber \\
&&+\frac{1}{2\sqrt{2}s}\,{\rm e}^\phi
\left(s{\cal F}^{(1)}+\gamma_5{\cal F}^{(2)}\right)
\gamma_\mu\epsilon
+\frac{1}{4\sqrt{2}s}\,{\rm e}^{-\phi} 
\left(s\,g_{(1)}+g_{(2)}\gamma_5\right)\gamma_\mu\epsilon\, ,
\eea
with the Cartan metric $K^{(\sigma)}_{ab}$ from  (\ref{c9a}).

To recapitulate, Eqs. (\ref{d22}), (\ref{d24}), (\ref{d29}), (\ref{d30}),
and  (\ref{d32}) provide an equivalent representation
of the ten-dimensional SUSY variations 
$\delta\hat{\chi}$ and $\delta\hat{\psi}_M$, 
as no truncation 
of the fermionic degrees of freedom has been done so far. 
Let us now assume that the parameter
${\epsilon}$ does not depend on the internal coordinates 
\be                                                             \label{d33}
\frac{\partial}{\partial z^{\rm m}}\, \epsilon=0\, .
\ee
This is consistent due to the appearance of the partial derivative  
in (\ref{d29}) discussed above, 
and eventually due to the fact that the internal
space is a group manifold. In the case of 
dimensional reduction on general homogeneous spaces
the dependence of spinors on internal coordinates 
is usually more complicated and is given 
in terms of Killing spinors on the internal
space \cite{Duff86}.

Let us now suppose  that 
$\delta\chi=\delta\psi_\mu=0$.
Eq. (\ref{d22}) then implies  that $\delta\hat{\chi}=0$,
Eq. (\ref{d30}) shows that $\delta\hat{\psi}_\mu=0$,
while Eq. (\ref{d29}) ensures in view of (\ref{d33})  that  
$\delta\hat{\psi}_{\rm m}=0$. 
As a result, all components of the ten-dimensional 
SUSY variations vanish, $\delta\hat{\chi}=\delta\hat{\psi}_M=0$. 
This shows that we can restrict our considerations to
the four-dimensional SUSY variations 
$\delta\chi$ and $\delta\psi_\mu$ given by 
(\ref{d24}) and (\ref{d32}).
The vanishing of these implies that the 
background bosonic configuration is  supersymmetric 
when lifted to ten dimensions.

We have completed our program of deriving the four-dimensional
theory from the ten-dimensional one.
Summarizing, in addition
to the  bosonic Lagrangian (\ref{c26}) we now have also
the four-dimensional supersymmetry transformations 
(\ref{d24}), (\ref{d32}). 
For $s=i$ these exactly coincide with the Lagrangian
and linearized SUSY variations of the Freedman-Schwarz model 
described in 
\cite{Freedman78}, up to a change of the overall sign of the metric:
\be                                     \label{d34}
g_{\mu\nu}\to -g_{\mu\nu},\ \ \
\gamma^\mu\to i\gamma^\mu,\ \  \ 
\gamma_\mu\to -i\gamma_\mu,\ \ \
 \gamma_5\to \gamma_5\,  .
\ee 
For $s=1$ Eqs. (\ref{c26}), (\ref{d24}), (\ref{d32}) give us
the Lagrangian
and SUSY variations of the Euclidean Freedman-Schwarz model.
This appears to be the N=4 gauged SU(2)$\times$SU(1,1) supergravity
in four-dimensional Euclidean space. Here N=4 is due to the fact that
the spinor supersymmetry parameter $\epsilon$ in (\ref{d24}), (\ref{d32})
is a multiplet of four Majorana spinors. 

Having obtained the theory, we shall now proceed with studying its vacuum
structure. Since we have a gauged supergravity, we shall mainly be
interested in solutions with non-Abelian gauge fields. 
In particular, we still need to explain the relation $g_{(2)}\to ig_{(2)}$
between the two models, as this apparently does not hold at the level 
of the full D=4 theories, which are rather  related via $s\to is$.

\section{Vacua}
\setcounter{equation}{0}

A supersymmetric vacuum is an on-shell 
bosonic configuration which is invariant under
some or all of the SUSY transformations. This 
manifests in the existence of non-trivial spinor parameters
$\epsilon$ for which the fermion SUSY variations vanish.
Such $\epsilon$'s are called supersymmetry Killing spinors. 
In an N=4 supergravity a vacuum can have at most 16 Killing spinors,
which is the number of the real components of $\epsilon$, and such 
a vacuum is called maximally supersymmetric. 

The Freedman-Schwarz model has no maximally supersymmetric vacua.
This is because the latter are expected to respect the maximal number of
spacetime isometries, while the model  
does not admit solutions with maximal symmetry.
In view of the relation to ten dimensional supergravity,
this is guaranteed by the  
``ten into four won't go" theorem: N=1, D=10
supergravity does not admit solutions of the form $M\times S^3\times S^3$,
where $M$ is a maximal symmetry space 
(Minkowski, de Sitter, or $AdS$) \cite{Freedman83}. 
One can also see directly that Eqs. (\ref{c19})--(\ref{c24}) for $s=i$
do not admit maximally symmetric solutions, since the dilaton 
potential in (\ref{c20}) has no stationary points. 
However, there are vacua in the model which are 
of the type of a direct product of the two
maximally symmetric spaces, $AdS_2\times E^2$, 
and these solutions preserve
half of the  supersymmetries \cite{Freedman84}.
The model also admits  domain-wall-type vacua with
half of the  supersymmetries preserved \cite{Cowdall97,Singh98},
as well as other vacuum solutions which can be of  various types and typically
preserve less then half of the supersymmetries \cite{Singh98,Klemm98}. 
When lifted to ten dimensions, some of the known FS vacua can be 
interpreted as the near-horizon geometries of certain intersecting
brane solutions \cite{Cowdall98,Singh98a}. Almost all known FS
vacua are characterized by the gauge fields belonging to the Cartan
subalgebra of the Lie algebra of SU(2)$\times$SU(2). Only
one solution is known 
whose gauge field is truly non-Abelian, this
is of  regular monopole type
\cite{Chamseddine97,Chamseddine98}, it
preserves 1/4
of the supersymmetries and will be briefly discussed below.

Let us now turn to the Euclidean Freedman-Schwarz model. In this 
case we can  get rid of the dilaton potential 
by choosing $g_{(1)}=g_{(2)}$. 
This allows us to set in 
Eqs. (\ref{c19})--(\ref{c24}) the vector fields to zero 
and scalar fields to constant values. The non-trivial 
bosonic equations 
then reduce to
\be                            \label{a-}
R_{\mu\nu}=0\, ,
\ee
whose solution can be any gravitational instanton ${\cal M}_4$. 
The conditions $\delta\chi=\delta\psi_\mu=0$ read
\bea                                            \label{a1}
\nabla_\mu\,\epsilon&=&0\, , \nonumber \\
g_{(1)}\, (1-\gamma_5)\,\epsilon&=&0\, ,
\eea 
with $\nabla_\mu$ being the geometrical covariant derivative,
which requires  
that ${\cal M}_4$ should admit chiral 
geometrical Killing spinors
(remember that the Euclidean Majorana condition
for $\epsilon$ 
is compatible with the Weyl condition). This
gives the simplest vacua in the EFS model, and these
can be uplifted  to D=10
with the use of Eqs. (\ref{c7}) and (\ref{c18})
leading to solutions of the form
${\cal M}_4\times S^3\times AdS_3$, 
the radii of the internal manifolds
being $1/g_{(1)}$. The D=10 dilaton is constant, while 
the antisymmetric tensor field coincides up to $g_{(1)}^2$ 
with the direct sum of the volume forms on the internal three-spaces. 

The EFS vacua described above include the flat instanton $E^4$,
which is maximally symmetric. In view of the chirality 
condition in (\ref{a1}) this has only eight Killing spinors and thus 
is not maximally supersymmetric. The reason for this is 
clear from the ten-dimensional point of view, since 
$E^4\times S^3\times AdS_3$ is not  maximally symmetric. 
For $g_{(1)}\to 0$ the second condition in 
(\ref{a1}) disappears and the number of supersymmetries doubles,
while the ten-dimensional configuration reduces to the flat metric.

One can also study more complex EFS vacua, in particular those 
with non-trivial scalars and  gauge fields. 
Such solutions are probably relatively easy to obtain
in the case where the gauge fields belong to the Cartan subalgebra. 
However, since we have a gauged supergravity, our primary interest
will be in configurations with essentially non-Abelian structures, 
and we shall  explicitly present such solutions below. 
Our strategy will be as follows. First, we shall further 
reduce the theory from D=4 to D=3
and recover in this case the relation between the FS and EFS models via
$g_{(2)}\to i g_{(2)}$, which has been the main motivation for
the present work. 
Next, we shall  impose spherical symmetry and  give a complete
derivation of  the supersymmetry constraints and the 
Bogomol'nyi equations. 
Finally we shall describe the known non-Abelian solutions 
of the Bogomol'nyi equations and their interpretation.

\subsection{Reduction to D=3}
We wish to further reduce the four-dimensional theory specified
by (\ref{c26}), (\ref{d24}), (\ref{d32}) to D=3.  
In order to  maintain consistency of the procedure 
we shall again work at the level of equations of motion and, 
as usual, keep both values of the parameter $s$. 
In brief, our dimensional reduction ansatz is
\be                                                     \label{a0}
\frac{\partial}{\partial x^0}=0,\ \ \ 
A^{(2)a}_\mu=0,\ \ \
A^{(1)a}_0=0,\ \ \ 
{\bf a}=0.
\ee
The first condition here means that 
$\partial/\partial x^0$ is a Killing vector,
and thus there is a gauge where 
all variables depend only on the spatial coordinates $x^{\ui}$
(we shall denote the spatial base space and tangent space 
indices by bold-faced letters ${\ui,\uk}=1,2,3$
and ${\ua,\ub}=1,2,3$, respectively). Next, 
the bosonic field equations (\ref{c19})--(\ref{c24}) show that 
one can consistently set the second gauge
field to zero, $A^{(2)a}_\mu=0$. However, we   
keep at the same time $g_{(2)}\neq 0$.
This leaves us with only one gauge field, 
$A^{a}_\mu\equiv A^{(1)a}_\mu$,  whose gauge group is SU(2). 
Without loss of generality we can assume that $g_{(1)}=1$. 
Next,  we require that $A^a_0=0$, and this
implies that the 
$\ast\! FF$ invariant of the gauge field vanishes.
It follows then 
from  Eq. (\ref{c24}) that one can consistently set the axion to zero,
${\bf a}=0$. Finally, we assume that the Killing vector 
$\partial/\partial x^0$ is 
hypersurface orthogonal, in which case the metric can be chosen as
\be                                                 \label{a2}
ds^2=s^2{\rm e}^{2V}dt^2+
h_{\ui\uk}\,dx^{\ui} dx^{\uk}\, .
\ee
As a result, we are now left with only the 
metric amplitude $V$, the three-metric 
$h_{\ui\uk}$, the dilaton $\phi$, and 
the three-dimensional gauge field $A^a_{\ui}$.
In the Lorentzian case $(s=i)$ our truncation corresponds
to the static and purely magnetic sector of the 
FS model with the second gauge
field set to zero. In the Euclidean $(s=1)$ domain the notions 
`static' and 'purely magnetic' do not have an invariant meaning. 
It will be convenient to use together with the three-metric
$h_{\ui\uk}$ also its conformally rescaled version, $g_{\ui\uk}$, 
the `spatial' line element being
\be                                                 \label{a1:0}
dl^2=h_{\ui\uk}dx^{\ui}dx^{\uk}=
{\rm e}^{-2V}g_{\ui\uk}dx^{\ui}dx^{\uk}=\delta_{\ua\ub}
\theta^{\ua}\theta^{\ub}\, ,
\ee
where $\theta^{\ua}$ are the basis one-forms, which are
dual  to the triad $e_{\ua}$.

One can verify that under the conditions in (\ref{a0}) and (\ref{a2})
the D=4 field equations (\ref{c19})--(\ref{c24}) consistently reduce to 
the Lagrangian equations for the action
\be                                                        \label{a3}
S_3=\int{\cal L}_3\sqrt{g_3}\, d^3x\, ,
\ee
with $g_3\equiv {\rm det}(g_{\ui\uk})$ and 
\bea
{\cal L}_3&=&\frac14\, \stackrel{(3)}{R}
-\frac12\,\partial_{\ui}\phi \,\partial^{\ui}\phi
- \frac12\,\partial_{\ui} V\,\partial^{\ui} V   \nonumber  \\
&-&\frac14\,
{\rm e}^{2\phi+2V}
F^{a}_{\ui\uk}F^{a\ui\uk} 
+\frac18\,
(1-\xi^2)\,{\rm e}^{-2\phi-2V}\, .      \label{a4}   
\eea
Here $F_{\ui\uk}=\partial_{\ui} A_{\uk}-
\partial_{\uk} A_{\ui}
+\varepsilon_{abc}A^b_{\ui}
A^c_{\uk}$ is the gauge field tensor,
$\stackrel{(3)}{R}$ is the Ricci-scalar for $g_{\ui\uk}$, 
and $\xi\equiv g_{(2)}/s$. Notice the following important fact:
the reduction above has been done for both values of $s$. At the same time, 
the $s$-dependence is now almost completely gone, and  
if we ignored the dilaton potential then the field equations would
be exactly the same in the Euclidean and Lorentzian cases. This is because
the system is `static' and `purely magnetic'.  It is only the explicit
dependence of the dilaton potential on $s$ that breaks the 
complete symmetry between the two cases. 

The action 
(\ref{a3}) admits the global symmetry
\be                                                                     
\label{a5}
\phi\to\phi+a,\ \ \ 
V\to V-a\, ,
\ee
and this implies that there is the Noether current, whose conservation law
reads
\be                                                                     
\label{a6}
\stackrel{(3)}{\nabla}\!_{\ui}\stackrel{(3)}{\nabla}\!^{\ui}(\phi-V)=0\, ,
\ee
where $\stackrel{(3)}{\nabla}$ is the covariant derivative with respect to 
$g_{\ui\uk}$. As a result,  we can consistently set
\be                                                                     
\label{a7}
V=\phi-\phi_\infty\, ,
\ee
where $\phi_\infty$ is the value of the dilaton at `spatial' infinity. 

Let us now consider the reduction of the D=4 SUSY variations in 
(\ref{d24}) and (\ref{d32}) to D=3. Inserting (\ref{a0}), (\ref{a2}) into 
(\ref{d24}) and (\ref{d32}) the result is 
\bea                                            \label{a8}
&&\delta\chi =\left(\frac{1}{\sqrt{2}}\,
\gamma^{\ui}\partial_{\ui}\phi\,
-\frac{1}{2}\,{\rm e}^\phi\,
\T_{a}F^a_{\ui\uk}\gamma^{\ui}\gamma^{\uk}
+\frac{1}{4}\,{\rm e}^{-\phi} (1-\gamma_5\xi)\right)\epsilon\, ,\nonumber \\
&&\delta\psi_{\uj}={\cal D}_{\uj}\epsilon
+\frac{1}{2\sqrt{2}}\,\left(-{\rm e}^\phi\,
\T_{a}F^a_{\ui\uk}\gamma^{\ui}\gamma^{\uk}
+\frac{1}{2}\,{\rm e}^{-\phi} 
(1+\gamma_5\xi)\right)\gamma_{\uj}\epsilon\,     .
\eea
Here the covariant derivative ${\cal D}_{\uj}\equiv\partial_{\uj}+
\frac14\,\omega_{\ua\ub,\uj}\gamma^{\ua}\gamma^{\ua}
+\T^{a}A_{\uj}^{a}$, where $\omega_{\ua\ub,\uj}$
is the spin-connection for $\theta^{\ua}$, and 
$[\T_a,\T_b]=\varepsilon_{abc}\T_c$ are the SU(2) generators. 
The $\gamma^{\ua}$'s are the four-dimensional 
gamma matrices for the spatial 
values of the index, $\gamma^\alpha=(\gamma^0,\gamma^{\ua})$, one has
$\gamma^{\ua}\gamma^{\ub}+\gamma^{\ub}\gamma^{\ua}
=2\delta^{\ua\ub}$ and
$\gamma^{\ui}=e^{\ \ui}_{\ua}\gamma^{\ua}$. 

The temporal component $\delta\psi_0$ obeys
\be                            \label{a9}
\delta\psi_0-\frac{1}{\sqrt{2}}\,\gamma_0\,\delta\chi=
\left.\left.\frac12\,\right(
\gamma_0\gamma^{\uk}\partial_{\uk}(V-\phi)\right)\epsilon=0\, ,
\ee
where the last equality on the right is due to (\ref{a7}). 
This shows that $\delta\psi_0$ 
is not independent and vanishes whenever other SUSY variations  vanish,
provided that the condition (\ref{a7}) holds, the latter thus being
one of the supersymmetry conditions. 

We have completed the reduction to D=3. 
The bosonic sector of the resulting theory is described by 
Eqs. (\ref{a3}), (\ref{a4}) together with the constraint in (\ref{a7}),
while the fermion SUSY transformations are given by (\ref{a8}). 
We now make the following  observation. 
All expressions above depend on $s$ only via the ratio $\xi=g_{(2)}/s$. 
For real values of $\xi$ we obtain the EFS model, while choosing
$\xi$  imaginary gives the FS model. This shows that starting from
the static, purely magnetic sector of the Freedman-Schwarz supergravity
and making the formal replacement $g_{(2)}\to i\, g_{(2)}$
gives the `static' and `purely magnetic' 
sector of the Euclidean Freedman-Schwarz theory. 
This explains the empirical observation made in Ref.\cite{Volkov99}
that a formal analytic continuation in the supergravity equations
gives a meaningful result -- because we obtain in this way another
consistent supergravity model. 

The D=3 field equations in (\ref{a3}), (\ref{a4}),  (\ref{a7}) allow one
to study the static  solutions.
As there is little
hope to directly solve the equations for the bosonic action (\ref{a3}),
one can start from the equations for the supersymmetry Killing spinors 
$\epsilon$ obtained from (\ref{a7}) by setting 
$\delta\chi=\delta\psi_{\uj}=0$.
The consistency conditions for these equations can be formulated as a 
set of first order Bogomol'nyi equations for the underlying bosonic
configuration. The Bogomol'nyi equations are compatible with the 
second order field equations and their solutions automatically
give supersymmetric vacua. So far, however, the corresponding 
construction has been carried out only 
for the spherically symmetric fields. 
These will be considered below.

\subsection{Spherical symmetry}

Let us consider the reduction of the 
D=3 theory described by Eqs. (\ref{a3}), (\ref{a4}) and  (\ref{a7}) 
to the spherically symmetric sector. 
The most general spherically symmetric 3-metric is
\be                                          \label{ha1}
dl^2={\rm e}^{2\lambda}\,dr^2+
{\rm e}^{2\mui}\,d\Omega^2\, ,
\ee
where $d\Omega^2=d\theta^2+\sin^2\theta\,d\varphi ^{2}$ is the 
line element of the unit sphere.
The spherically symmetric and purely magnetic 
Yang-Mills field  is given by 
\be                                                     \label{aa2}
A=w\ (-\T_2\,d\theta +\T_1\,
\sin \theta \,d\varphi )+\T_3\,\cos \theta \,d\varphi .
\ee
Here  $\lambda$, $\mui$, $w$ as well as the dilaton $\phi$
depend only on the radial coordinate $r$, and the 
reparameterization invariance $r\to\tilde{r}(r)$ implies that 
one coordinate condition can be imposed on the four amplitudes.
Taking the condition in (\ref{a7}) into account, the 4-metric reads
\be                                          \label{aa1a}
ds^2=s^2{\rm e}^{2(\phi-\phi_\infty)}\,dt^2+dl^2\, .
\ee
The complete set of the field
equations for the bosonic action (\ref{a3}) is
\bea                                                          
{\rm e}^{2\mui}\,(\mui'^2+2\mui'\phi')
&=&{\rm e}^{2\mui}\,\phi'^2+2{\rm e}^{2\phi}\,w'^2+
{\rm e}^{2\lambda}\left(1-{\rm e}^{2\phi-2\mui}\,
(w^2-1)^2\right)                 \nonumber     \\
&+& \frac14\, (1-\xi^2)\,{\rm e}^{2\lambda-2\phi+2\mui}\, ,     \nonumber  
   \\
\left({\rm e}^{\phi-\lambda+2\mui}\,\mui'\right)^\prime&=&
{\rm e}^{\phi+\lambda}\left(1-{\rm e}^{2\phi-2\mui}\,
(w^2-1)^2\right)+ \frac14\, (1-\xi^2)\,{\rm e}^{\lambda-\phi+2\mui}\, ,        
                            \nonumber      \\
\left({\rm e}^{\phi-\lambda+2\mui}\,\phi'\right)^\prime&=&
2{\rm e}^{3\phi-\lambda}\,w'^2
+{\rm e}^{3\phi+\lambda-2\mui}\,(w^2-1)^2
+ \frac14\, (1-\xi^2)\,{\rm e}^{\lambda-\phi+2\mui}\, ,   \nonumber \\
\left({\rm e}^{3\phi-\lambda}\, w'\right)^\prime&=&
{\rm e}^{3\phi+\lambda-2\mui}\, w(w^2-1) \,   ,      \label{aa7}
\eea
with $^\prime :=\frac{d}{dr}$. The same 
equations can be obtained by varying the four dimensional 
action (\ref{c26}) and using Eq. (\ref{a7}).

The supersymmetric vacua that we shall be considering 
are solutions to these equations for which there are  
 non-trivial $\epsilon$'s such that $\delta\chi=\delta\psi_{\uj}=0$. 
Of course, it is very difficult to directly solve the non-linear 
equations in (\ref{aa7}) (apart from some trivial cases).
For this reason we shall start
from the equations $\delta\chi=\delta\psi_{\uj}=0$
for the Killing spinors $\epsilon$. These equations
are sometimes called supersymmetry constraints, and they are 
generically inconsistent. One can analyze the consistency conditions
under which non-trivial solutions for  the ${\epsilon}$'s
exist. These conditions can be given in the form of a set
of nonlinear first order differential equations for the
underlying bosonic configuration -- usually called Bogomol'nyi equations.
The Bogomol'nyi equations are compatible with the second order
field equations, and their solutions therefore describe 
supersymmetric vacua. 

Following this strategy, our procedure will be to analyze the 
supersymmetry constraints $\delta\chi=\delta\psi_{\uj}=0$ obtained 
from Eqs. (\ref{a8}) in the case of spherical symmetry. 

\subsubsection{The supersymmetry constraints} It is convenient to 
choose the isotropic gauge in the line element (\ref{aa1a}), 
$r{\rm e}^\lambda={\rm e}^\mui$, and then to pass
to the Cartesian coordinates $x^{\ui}$ with
$r=\sqrt{\delta_{\ui\uk}x^{\ui}x^{\uk}}$. 
The  three-metric reads
\be                                                 \label{a11}
dl^2=
{\rm e}^{2\lambda}\,(dr^2+
r^2\,d\Omega^2)=
{\rm e}^{2\lambda}\delta_{\ui\uk}dx^{\ui}dx^{\uk}\, .
\ee
The triad vectors and one-forms are 
$e_{\ua}={\rm e}^{-\lambda}\partial_{\ua}$
and
$\theta^{\ua}={\rm e}^{\lambda}dx^{\ua}$, 
respectively. The spin-connection is
\be                                           \label{a12}
\omega_{\ua\ub,\uc}=\lambda'\, {\rm e}^{-\lambda}
(n_{\ub}\delta_{\ua\uc}-n_{\ua}\delta_{\ub\uc}).
\ee
Here and below $x^{\ua}\equiv \delta^{\ua}_{\uk}x^{\uk}$, 
$\partial_{\ua}\equiv\partial/\partial{x^{\ua}}$, and
$n^{\ua}\equiv x^{\ua}/r$.
The triad components of the gauge field (\ref{aa2}) read
\be                                \label{a12a}
A^a_{\ua}={\rm e}^{-\lambda}\,\frac{1-w}{r}\, 
\varepsilon_{a\ua\ub}\,n^{\ub}\, ,
\ee
and the gauge field strength is
\be                                      \label{a13}
F^a_{\ua\ub}={\rm e}^{-2\lambda}\varepsilon_{\ua\ub\uc}\,(
{\rm f}_1\,\delta^{a\uc}+{\rm f}_2\,  n^a n^{\uc})\, ,
\ee
with
\be
{\rm f}_1=\frac{w'}{r}\, ,\ \ \ \ \ 
{\rm f}_2=\frac{w^2-1}{r^2}-\frac{w'}{r}\, .
\ee
Here we handle
the triad and internal indices with 
the metric tensors $\delta_{\ua\ub}$ and $\delta_{ab}$
 and allow for objects with mixed indices
like $\varepsilon_{a\ua\ub}$ and $\delta_{a\ua}$.

Let us now introduce four different spinor two-spaces and denote
the corresponding Pauli matrices  by $\underline{\sigma}^{\ua}$, 
${\sigma}^{\ua}$, ${\tau}^{\ua}$, $\underline{\tau}^{\ua}$, 
respectively. We choose the gamma matrices
and  group generators as
\be                                      \label{a14}
\gamma^{\ua}=\underline{\sigma}^2\otimes\sigma^{\ua},\ \ \
\gamma_5=\underline{\sigma}^3\otimes\1_2\, ,\ \ \
\T^a=\frac{1}{2i}\,\tau^a\otimes \underline{\1}_2\, .
\ee
Here $\underline{\1}_2$ acts in the $\underline{\tau}^{\ua}$ space.
In what follows we shall not write down explicitly 
the direct product sign and the unit operators.
Taking into account all the  definitions above,
the SUSY variations in (\ref{a8}) assume the form
\bea
\delta\chi&=&\frac{1}{\sqrt{2}}\,{\rm e}^{-\lambda}\phi'\,
\underline{\sigma}^2\!
\left(\vec{n}\vec{\sigma}\right)\epsilon-
\left.\left.
\frac12\,{\rm e}^{\phi-2\lambda}\, \right(
{\rm f}_1\left(\vec{\sigma}\vec{\tau}\right)+
{\rm f}_2
\left(\vec{n}\vec{\sigma}\right)\left(\vec{n}\vec{\tau}\right)\right)\epsilon
\nonumber  \\
&+&\frac14\,{\rm e}^{-\phi}
(1-\xi\,\underline{\sigma}^3)\,\epsilon \, , \nonumber 
\eea
\bea
\delta\psi_{\ua}&=&{\rm e}^{-\lambda}
\left(\partial_{\ua}+
\frac{i}{2}\,\lambda'\,\varepsilon_{\ua\ub\uc}\,n^{\ub}\sigma^{\uc}
+\frac{i}{2}\frac{w-1}{r}\,\varepsilon_{{\ua\ub}c}\,n^{\ub}\tau^{c}
\right)\epsilon    \nonumber \\
&-&\left.\left.
\frac{1}{2\sqrt{2}}\,{\rm e}^{\phi-2\lambda}
\underline{\sigma}^2\right(
{\rm f}_1\left(\vec{\sigma}\vec{\tau}\right)+
{\rm f}_2
\left(\vec{n}\vec{\sigma}\right)\left(\vec{n}\vec{\tau}\right)\right)
\sigma_{\ua}\,\epsilon  \nonumber  \\
&+&\frac{1}{4\sqrt{2}}\,{\rm e}^{-\phi}
(\underline{\sigma}^2
-i\xi\,\underline{\sigma}^1)\,\sigma_{\ua}\,\epsilon \, .\label{a16}
\eea
Here $(\vec{n}\vec{\sigma})=\delta_{\ua\ub}n^{\ua}\sigma^{\ub}$,
also 
$(\vec{n}\vec{\tau})=\delta_{{\ua}b}n^{\ua}\tau^{b}$,
and 
$(\vec{\sigma}\vec{\tau})=\delta_{{\ua}b}\sigma^{\ua}\tau^{b}$.

Let us recall that $\epsilon$ is a 16-component spinor subject 
to the Majorana condition
\be                                     \label{a17:0}
(\epsilon)^\ast=\sigma^2\,\tau^2
\,\underline{\tau}^1\,
\epsilon
\ee
for real $\xi$ (Euclidean theory), and
\be                                     \label{a17}
(\epsilon)^\ast=\underline{\sigma}^2\,\sigma^2\,\tau^2
\,\underline{\tau}^2\,
\epsilon
\ee
for imaginary $\xi$ (Lorentzian theory). 
Now, since  we are considering spherically symmetric 
backgrounds, it is natural to choose  $\epsilon$ to be 
an eigenstate of the total angular momentum. Let us first study 
the sector with zero angular
momentum, $J=0$; the case of $J>0$ will be discussed later.  
We choose 
\be                                     \label{a18}
\epsilon=\epsilon_q\equiv\left.\left.
\left(\Psi^{(+)}_q\,
\underline{\psi}_q
+\Psi^{(-)}_q\, \underline{\sigma}^2\underline{\psi}_q
(\vec{n}\vec{\sigma})\right)
\right(\psi_{+}\,\chi_{-}-\psi_{-}\,\chi_{+}\right)
\underline{\chi}\, .
\ee
Here $q=\pm 1$,  $\Psi^{(\pm)}_q$ are functions of $r$,
while
$\underline{\psi}_q$, $\psi_\pm$, $\chi_\pm$, and $\underline{\chi}$
are constant two-component spinors from the four different 
spinor spaces in which the operators $\underline{\sigma}^{\ua}$, 
${\sigma}^{\ua}$, ${\tau}^{\ua}$, $\underline{\tau}^{\ua}$, 
respectively, act. 
One has $\underline{\sigma}^{3}\underline{\psi}_q
=q\underline{\psi}_q$, 
${\sigma}^{3}\psi_{\pm}=\pm\psi_{\pm}$, and 
$\tau^3\chi_{\pm}=\pm\chi_{\pm}$. 
Notice that the ansatz for $\epsilon$ in (\ref{a18}) 
is the most general expression annihilated by the total 
(orbital plus spin plus isospin) angular momentum operator,
\be                                             \label{a19}
\left(-i\varepsilon^{\ua\ub\uc}x_{\ub}\partial_{\uc}+
\frac12\,\sigma^{\ua}+\frac12\,\tau^{\ua}\right)\epsilon=0\, .
\ee
Let us insert the ansatz (\ref{a18}) into (\ref{a16}) and set 
the left-hand sides to zero. After some spinor algebra the angular
dependence decouples and we obtain a system of equations for 
${\Psi}^{(\pm)}_q$: 
\bea                                   \label{a23}
\left(\frac{d}{dr}-\frac{\phi'}{2}\right)
\Psi^{(\pm)}_q
\pm\sqrt{2}{\rm e}^{\phi-\lambda}\,
\frac{w'}{r}\Psi^{(\mp)}_q=0\, ,  \\                            
\left(\frac{\lambda'}{2}+\frac{1\mp w}{2r}\right) \label{a24}
\Psi^{(\pm)}_q
+\left(\frac{{\rm e}^{\phi-\lambda}}{2\sqrt{2}}\,
\frac{1-w^2}{r^2}
+\frac{{\rm e}^{\lambda-\phi}}{4\sqrt{2}}(1\pm q\xi)
\right)\Psi^{(\mp)}_q=0\, ,    \\                           
\frac{\phi'}{\sqrt{2}}\,           \label{a25}     
\Psi^{(\pm)}_q
+\left({\rm e}^{\phi-\lambda}(
\frac{w^2-1}{2r^2}\mp\frac{w'}{r})
+\frac{{\rm e}^{\lambda-\phi}}{4}(1\pm q\xi)
\right)\Psi^{(\mp)}_q=0\, .
\eea
These are the supersymmetry constraints we are
interested in. 

\subsubsection{The Bogomol'nyi equations}
It is not difficult to find the consistency conditions for 
Eqs. (\ref{a23})--(\ref{a25}). Consider, for example, 
Eq. (\ref{a25}). This is in fact a system of two homogeneous algebraic 
equations, which has a non-trivial solution only if the determinant
of the coefficient matrix vanishes. The same is true with (\ref{a24}). 
In addition, the solutions obtained from (\ref{a24}) and (\ref{a25})
should agree.
As a result, we obtain three algebraic conditions for the 
coefficients in Eqs. (\ref{a24}) and (\ref{a25}), which  
can be expressed in the form
\bea
1+r\frac{d\lambda}{dr}&=&
\sqrt{w^2+\frac{1}{8}\,
 {\rm e}^{2(\lambda+\ln r-\phi)}\,
\left((B-1)^2-q^2\xi^2\right)}\equiv\nu \, ,  \nonumber
\\
Ar\frac{dw}{dr}&=&
2q\xi\,w\nu+q^2\xi^2(w^2-1)-2w^2(B+1)\, , \nonumber
\\
Ar\frac{d\phi}{dr}&=&-(B+1)
(q\xi\nu+w(B-1))\, ,                   \label{a26}
\eea
with 
$A\equiv 8w\nu\,{\rm e}^{2(\phi-\lambda-\ln r)}
+q\xi\,(B-1)$ and $B\equiv 2{\rm e}^{2(\phi-\lambda-\ln r)}(w^2-1)$.  
Under these conditions Eqs. (\ref{a24}) and (\ref{a25}) 
specify the algebraic relation between $\Psi^{(+)}_q$ and
$\Psi^{(-)}_q$.  Next, this relation should be consistent with the 
remaining differential constraints in (\ref{a23}). A direct
calculation shows that the required consistency holds by virtue
of Eqs. (\ref{a26}).
The differential constraints (\ref{a23}) then fix
the solution for $\Psi^{(\pm)}_q$ 
uniquely --
up to a normalization constant,
\be                                  \label{spinor}
\Psi^{(+)}_q+i\Psi^{(-)}_q=\exp\left(i\sqrt{2}
\int^{r}_{r_0}{\rm e}^{\phi-\lambda}\,\frac{w'}{r}\,dr\right)\, .
\ee 
As a result, the equations in 
(\ref{a26}) constitute a complete set of 
consistency conditions under which the supersymmetry 
Killing spinors exist.  
These first order Bogomol'nyi equations 
are compatible (for $q=\pm 1$) with the 
second order field equations in (\ref{aa7}). 
In orther words, they 
are first integrals for the field equations. 
Although this fact is expected, its direct 
verification is not at all trivial and 
provides a very good check of the consistency of the whole 
procedure. Any solution
of the  Bogomol'nyi equations hence fulfills the field equations 
and admits non-trivial supersymmetry Killing spinors, thus describing
a supersymmetric vacuum. 

Before we pass to studying the  Bogomol'nyi equations, let us count
the number of  supersymmetry 
Killing spinors obtained from (\ref{a18}), (\ref{a23})--(\ref{a25}). 
Consider first the EFS case, where $\xi$ is real. 
Then all coefficients in Eqs. (\ref{a23})--(\ref{a25}) are real and the 
solution for $\Psi^{(\pm)}_q$ in (\ref{spinor}) 
can be chosen to be real as well. 
As was mentioned above, this solution is specified uniquely. 
We do not obtain solutions
for both values of $q$, since 
the  Bogomol'nyi equations for $\xi\neq 0$ are not 
invariant under $q\to -q$, but only under 
\be
q\to -q,\ \ \ \ \ w\to -w\, .
\ee
Thus, unless $w\equiv 0$, the value of $q$ is fixed
by the background configuration and there is only one
solution for the  $\Psi^{(\pm)}_q$.
Notice however that Eq. (\ref{a18})
contains an additional degeneracy due to the arbitrary constant spinor
$\underline{\chi}$.
The Majorana condition imposes the restrictions
\be                                             \label{a20}
\underline{\psi}_q=\left(\underline{\psi}_q\right)^\ast,\ \ \ \ \
\underline{\tau}^1\underline{\chi}=-\left(\underline{\chi}\right)^\ast\, .
\ee
There are two independent solutions to these conditions for a 
given $q$.
This finally gives two supersymmetry Killing spinors. 
For $\xi=0$ the number of supersymmetries doubles, since 
the  Bogomol'nyi equations are then 
invariant under $q\to-q$ and both 
values of $q$ are allowed in (\ref{a18}) and (\ref{a20}). 

Consider now the FS case, where $\xi$ is imaginary. Then 
$\xi=0$ is the only allowed value, since otherwise the  Bogomol'nyi 
equations contain imaginary coefficients and their solutions are not real. 
The solution for
$\epsilon$ is given by (\ref{a18}), where one can take both values of 
$q$,
and again there is an additional degeneracy due to $\underline{\chi}$.
In order to fulfill the Majorana condition in this case one should take
linear combinations of solutions with different $q$. 
Omitting the index $q$ of $\underline{\psi}_q$, the 
Majorana condition in (\ref{a17}) reduces to
\be                                             \label{a21}
(\underline{\sigma}^2\underline{\psi})\,
(\underline{\tau}^2\underline{\chi})=
-\left(\underline{\psi}\,\underline{\chi}\right)^\ast,
\ee
and this has four independent solutions. 

To recapitulate, supersymmetric vacua exist in 
the EFS model for arbitrary real $\xi$,  and 
in the FS case 
for $\xi=0$ only. 
For $\xi\neq 0$ the vacua 
preserve two supersymmetries.
For $\xi=0$, when one of the two gauge coupling constants
vanishes and the theory is `half-gauged', 
the vacuum admits four supersymmetry Killing spinors
and fulfills the equations of both FS 
and EFS models. 

It is worth noting that the above analysis uses only the 
Killing spinors from the sector with zero angular momentum, $J=0$. 
At the same time, there could be additional Killing spinors
for $J>0$, even though the background is spherically symmetric.  
Indeed, as was mentioned above, 
the $E^4$ instanton solution, which is given by 
$\lambda=\phi=0$ and $w=\xi=1$, 
has eight Killing spinors. Of these only
two are recovered in the system  (\ref{a23})--(\ref{a25}), 
and the remaining six must therefore reside  in  sectors with $J>0$. 
These sectors therefore should also be taken into consideration.
Now, one can show that any additional 
Killing spinors can only exist for $J=1$.
In this  case the whole procedure described above can be
repeated, which leads to a system of 
seven non-linear Bogomol'nyi equations for the three amplitudes 
$w$, $\phi$, and $\lambda$. The $E^4$ instanton fulfills these
new equations, which accounts for its additional six Killing spinors.
However, since the equations are overdetermined,
it is unclear whether they admit any other solutions. 
We shall not concentrate on this  here but rather pass to
considering the solutions of 
the Bogomol'nyi equations (\ref{a26}). 
For these solutions one can show that all their
Killing spinors  live in the $J=0$ sector.  

\subsubsection{Vacuum solutions for $\xi=0$. Supersymmetric monopoles}
It is not difficult to see that 
the Bogomol'nyi equations in (\ref{a26}) 
can be made autonomous, in which case the system 
essentially 
reduces to one first order differential equation.
This equation turns out to be rather complicated,
but it can be analyzed 
for special values of $\xi$. 
Let us consider the value $\xi=0$, in which case the solutions 
will have N=1 supersymmetry.  
Denoting $x\equiv w^{2}$ and 
$R\equiv \frac{1}{2}\,e^{2(\lambda+\ln r-\phi) }$,
Eqs. (\ref{a26}) give 
\cite{Chamseddine97,Chamseddine98}
\be
x (R+x-1)\frac{dR}{dx}+(x+1)\,R+(x-1)^{2}=0\, .  \label{a27a}
\ee
The two remaining 
Bogomol'nyi equations are solved by quadratures
as soon as  the solution $R(x)$ is obtained. After  the substitution
\cite{Chamseddine97}
\be
x=\rho^{2}\,e^{y(\rho)},\ \ \ \ \ \ \
R=-\rho\frac{dy (\rho)}{d\rho}-\rho^{2}\,e^{y
(\rho)}-1,                                             \label{a28}
\ee
the Abel equation (\ref{a27a}) 
reduces to the Liouville equation
\be
\frac{d^{2}y}{d\rho^{2}}=2\, e^{y},  \label{a29}
\ee
which is integrable. This gives a globally
regular solution described by
\be
d{ s}^{2}=2\,{\rm e}^{2\phi}\left(
s^2dt^{2}+d\rho^{2}+R\,d\Omega^2\right) ,  \label{a30}
\ee
\be                                                   \label{a31}
R=2\rho\coth \rho-\frac{\rho^{2}}{\sinh ^{2}\rho}-1,\ \ 
w=\pm \frac{\rho}{\sinh \rho},\ \ 
{\rm e}^{2\phi }= \frac{\sinh \rho}{\sqrt{R}},   
\ee
where the  scaling symmetry (\ref{a5}) has been used to set
$\phi(0)=0$. The new result we obtain here as
compared to that of \cite{Chamseddine97} is that this solution
turns out to be the vacuum of both the FS and EFS models, preserving in 
each case 1/4 of the supersymmetries. Corresponding to this there is the 
parameter $s$ in the metric in (\ref{a30}). 

For $s=1$ the solution describes a globally regular
Riemannian manifold with an essentially non-Abelian gauge field.
Since the configuration does not depend on $t$, the action
is infinite. Passing to the Lorentzian sector via choosing $s=i$,
the solution describes a globally regular magnetic monopole
with unit charge. This is geodesically complete and 
globally hyperbolic \cite{Chamseddine97,Chamseddine98}. 
Unfortunately, the ADM mass is infinite and the solution
is not asymptotically flat -- due to the dilaton potential. 
In view of its supersymmetry, it is very plausible that the 
Lorentzian solution is stable, while for its Euclidean 
counterpart the notion of dynamical stability makes no sense. 

\begin{figure}
\vspace{3 mm}
\hbox to\hsize{
  \epsfig{file=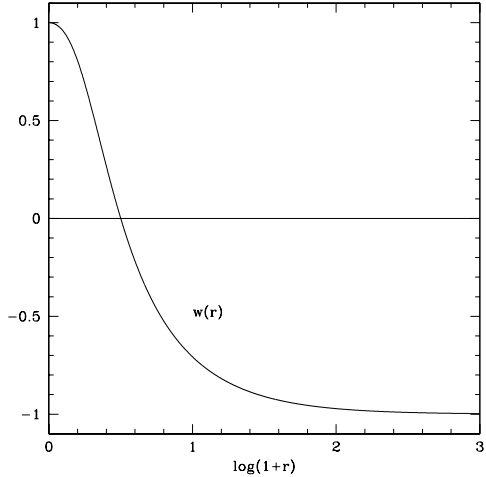,width=0.48\hsize,%
      bbllx=0.5cm,bblly=5cm,bburx=20cm,bbury=22cm}\hss
  }
\caption{The numerical solution of the Bogomol'nyi
equation (\ref{a33}).}
\label{fig1}
\vspace{3 mm}
\end{figure}

\subsubsection{Vacuum solutions for $\xi=1$. Supersymmetric sphalerons}
For any $\xi\neq 0$ solutions of the Bogomol'nyi equations (\ref{a26})
will preserve only 1/8 of the supersymmetries. Let us consider the value
$\xi=1$, in which case the dilaton potential vanishes. 
After some transformations 
described in \cite{Volkov99}
the Bogomol'nyi equations can be reduced to
\be                                              \label{a33}
\frac{1}{2r}\,\frac{dw}{dr}=\frac{1-w^2}{4r^2}
-\frac{(w+1)^3}{8}+\frac{(w-1)^3}{8r^4}\, ,
\ee
which is invariant under $r\to1/r$, $w\to-w$.
When  the solution $w(r)$ is found, the whole configuration is 
reconstructed as follows. Computing the combination
\be                                                  \label{a34}
U=\frac{r^2(1+w)^2+(1-w)^2}{r^2(1+w)^2-(1-w)^2}\, ,
\ee
the metric function 
$\lambda$ is obtained from 
\be                                                   \label{a35}
\lambda=
\ln(2)+\int_{0}^{r}\left(\frac{U+w}{2}-1\right)\frac{dr}{r},
\ee
while the dilaton is given by 
\be                                                  \label{a36}
\phi=
\lambda+\ln(r)+\frac12\,
\ln\left(\frac{(U+w)^2-2w^2-2}{2\,(w^2-1)^2}\right)\, ,
\ee
which is normalized such that $\phi(0)=0$. 
The metric is 
\be                                                      \label{a37}
ds^{2}={\rm e}^{2(\phi-\phi_\infty)}\, dt^{2}+{\rm e}^{2\lambda}
\left(dr^2+r^2d\Omega^2\right).
\ee
Unfortunately, analytical solutions to Eq. (\ref{a33}) are unknown 
(apart from singular ones \cite{Volkov99}). 
The numerical integration
(see Fig.\ref{fig1})
reveals the existence of a globally regular solution
in the interval $r\in[0,\infty)$ which monotonically interpolates 
between the values specified by the local asymptotic solutions:
$w=1-\frac23\,r^2+O(r^4)$ for $r\to 0$ and 
$w=-1+2\sqrt{2}\,r^{-1}+O(r^{-2})$ as $r\to\infty$. 
This gives a globally regular 
supersymmetric Euclidean solution with  non-trivial
Yang-Mills field and infinite action.

One can pass to  the Lorentzian sector  
by changing the sign of  $dt^2$  in the metric (\ref{a37}).
The resulting configuration fulfills the equations of 
motion of the Einstein-Yang-Mills-Dilaton (EYMD) model 
with the action
\be                                                       \label{a39}
S_{\rm EYMD}=
\int \left( \frac{1}{4}\,R-\frac{1}{2}\,\partial _{\mu }\phi \,\partial
^{\mu }\phi -\frac{1}{4}\,e^{2\phi }\,F_{\mu \nu }^{a}F^{a\mu \nu }
\right) \sqrt{-{g}}\,d^{4}x\, .  
\ee
The static and purely magnetic sector of this model 
can be embedded into the  
heterotic string theory. 
However, the solution is not supersymmetric in this model,
since it is not self-dual \cite{Strominger90}. 
Nevertheless,
this EYMD solution is interesting as it describes a regular
particle-like object with finite ADM mass $M$ 
determined by the asymptotic behaviour
of the metric in (\ref{a37}), 
${\rm e}^{2(\phi-\phi_\infty)}=1-2M{\rm e}^{-\lambda(\infty)}\,
r^{-1}+O(r^{-2})$ 
\cite{Volkov99}. 
This EYMD particle
is static, spherically symmetric and neutral, but it has a 
non-trivial purely magnetic
gauge field that asymptotically vanishes like $1/r^3$. 
It turns out that this solution resembles  the well-known
sphaleron solution of the Weinberg-Salam theory \cite{Manton83}. 
Specifically, one can show that the solution relates to the 
top of the potential barrier between the topological
vacua of the EYMD theory, which implies in particular that 
it is unstable. This suggests the name `EYMD sphaleron'.
One can argue that this solution is responsible for 
fermion number non-conserving processes in heterotic string theory. 
Despite its instability, it bears an imprint of supersymmetry
as it fulfills the first order Bogomol'nyi equations. Passing
back to the Euclidean theory, the counterpart of the EYMD 
sphaleron is a genuinely supersymmetric configuration, and we call it
`supersymmetric sphaleron'. 

Historically, it was the EYMD sphaleron
solution which was first obtained by numerical integration of the 
second order field equations for the action (\ref{a39}) 
\cite{Donets93,Lavrelashvili93,Bizon93a}. Only later it was 
discovered in \cite{Volkov99} that the solution fulfills the 
first order 
Bogomol'nyi equation (\ref{a33}), and it was conjectured 
that the configuration becomes supersymmetric upon continuation
to the Euclidean sector. The justification of this 
conjecture has been the main subject of the present paper.

\section{Conclusion}
In this paper we have studied the dimensional
reduction of the N=1, D=10 supergravity on 
$S^3\times AdS_3$. 
The resulting four-dimensional theory is Euclidean N=4 gauged
SU(2)$\times$SU(1,1) supergravity with the 
bosonic Lagrangian 
(\ref{c26}) and the fermion supersymmetry 
transformations specified by (\ref{d24}), (\ref{d32}).
An interesting feature of this model is that its dilaton potential U$(\phi)$
can be positive, negative, or zero, depending on the values of the 
two gauge coupling constants. This allows one to apply the model 
for generating various solutions with gravitating Yang-Mills fields 
via solving the Bogomol'nyi equations, which sometimes even
gives solutions in a closed analytical form. The two examples --
monopole-type and sphaleron-type non-Abelian vacua --
were described above. 
In view of the relation to N=1, D=10 supergravity, which is in turn 
related to D=11 supergravity \cite{Chamseddine81}, 
any solution of the theory can be uplifted to become 
a vacuum of string or M theory. 

It is worth noting that reductions on $AdS\times$Sphere 
are often considered; see for example \cite{Boonstra98}.
In particular, gravity and string theory on $AdS_3\times S^3$
have been studied in detail \cite{Behrndt98,Giveon98}.
In our analysis, however, the emphasis is quite different, 
since we are interested not in an effective   Lorentzian
theory with an $AdS$  ground state, but rather  
in a Euclidean theory admitting $E^4$ as a vacuum. 
Let us also note that in most cases the dimensional reduction 
is performed only at the level of the Lagrangian, which raises 
the issue of consistency; 
see \cite{Cvetic99,Lu99} for a recent discussion.
Our results on the other hand provide an example of a 
consistent reduction carried out at the level of equations of motion.

One can expect that the model described above admits  interesting 
solutions also beyond the static, spherically symmetric and purely
magnetic sector with U$(\phi)=0$. 
For U$(\phi)>0$  these could be, for example, compact instantons,
and possibly also  non-compact, asymptotically flat configurations. 
For U$(\phi)=0$ the theory probably admits non-compact
instantons with finite action. One can also study static 
multi-sphaleron solutions by deriving the Bogomol'nyi equations 
from Eqs. (\ref{a8}). 
An interpretation of the Euclidean solutions 
can sometimes be obtained by continuation
to the Lorentzian sector -- negative energy states 
will not arise if the SU(1,1) gauge field vanishes.
All solutions for gravitating Yang-Mills 
fields of this type are expected to be relevant
for string/M-theory 
and in the context of the general study of 
non-linear phenomena in field theory.

\section*{Acknowledgments} 
I am greatly indebted to Dieter Maison for numerous
communications and his moral support. 
I would like also to thank Andreas Wipf for some
discussions and Tom Heinzl for the careful reading 
of the manuscript.  


\begin{thebibliography}{10}

\bibitem{Duff95}
M.J. Duff, R.R. Khuri, and J.X. Lu.
\newblock {\ssl String solitons}.
\newblock {\em Phys.Rep.}, {\bf 259},\ 213--326, 1995.

\bibitem{Gauntlett93}
J.P. Gauntlett, J.A. Harvey, and J.T. Liu.
\newblock {\ssl Magnetic monopoles in string theory}.
\newblock {\em Nucl.Phys.}, {\bf B 409},\ 363--381, 1993.

\bibitem{Strominger90}
A.~Strominger.
\newblock {\ssl Heterotic solitons}.
\newblock {\em Nucl.Phys.}, {\bf B 343},\ 167--184, 1990.

\bibitem{Gibbons94a}
G.W. Gibbons, D.~Kastor, L.A.J. London, P.K. Townsend, and J.~Traschen.
\newblock {\ssl Supersymmetric self-gravitating solitons}.
\newblock {\em Nucl.Phys.}, {\bf B 416},\ 850--879, 1994.

\bibitem{Gibbons95a}
G.W. Gibbons and P.K. Townsend.
\newblock {\ssl Anti-gravitating BPS monopoles and dyons}.
\newblock {\em Phys.Lett.}, {\bf B 356},\ 472--278, 1995.

\bibitem{Nastase99}
H.~Nastase, D.~Vamar, and P.~van Nieuwenhuizen.
\newblock {\ssl Consistent nonlinear KK reduction of 11D supergravity on
  $AdS_7\times S_4$ and self-duality in odd dimensions}, 1999.
\newblock {\tt hep-th/9905075}.

\bibitem{Cvetic99}
M.~Cvetic, J.T. Liu, H.~L\"u, and C.N. Pope.
\newblock {\ssl Domain-wall supergravities from sphere reduction}, 1999.
\newblock {\tt hep-th/9905096}.

\bibitem{Lu99}
H.~L\"u and C.N. Pope.
\newblock {\ssl Exact embedding of N=1, D=7 gauged supergravity in D=11}, 1999.
\newblock {\tt hep-th/9906168}.

\bibitem{Volkov98}
M.S. Volkov and D.V. Gal'tsov.
\newblock {\ssl Gravitating Non-Abelian solitons and black holes with
  Yang-Mills fields}.
\newblock {\em Phys.Rep.}, {\bbf 319},\ 1--83, 1999.

\bibitem{Chamseddine97}
A.H. Chamseddine and M.S. Volkov.
\newblock {\ssl Non-Abelian BPS monopoles in N=4 gauged supergravity}.
\newblock {\em Phys.Rev.Lett.}, {\bbf 79},\ 3343--3346, 1997.

\bibitem{Chamseddine98}
A.H. Chamseddine and M.S. Volkov.
\newblock {\ssl Non-Abelian solitons in N=4 gauged supergravity and leading
  order string theory}.
\newblock {\em Phys.Rev.}, {\bbf D 57},\ 6242--6254, 1998.

\bibitem{Freedman78}
D.Z. Freedman and J.~Schwarz.
\newblock {\ssl N=4 supergravity model with local SU(2)$\times$SU(2)
  invariance}.
\newblock {\em Nucl.Phys.}, {\bf B 137},\ 333--339, 1978.

\bibitem{Gunaydin84}
M.~Gunaydin, G.~Sierra, and P.~K. Townsend.
\newblock {\ssl Vanishing potentials in gauged N=2 supergravity,\  
an application of jordan algebras}.
\newblock {\em Phys.Lett.}, {\bf B 144},\ 41--53, 1984.

\bibitem{Volkov99}
M.S. Volkov and D.~Maison.
\newblock {\ssl Bogomol'nyi equations for Einstein-Yang-Mills theory}.
\newblock {\tt hep-th/9904174}, to appear in {\it Nucl.Phys.}

\bibitem{Scherk79}
J.~Scherk and J.~Schwarz.
\newblock {\ssl How to get masses from extra dimensions}.
\newblock {\em Nucl.Phys.}, {\bf B 153},\ 61--88, 1979.

\bibitem{Chamseddine81}
A.H. Chamseddine.
\newblock {\ssl N=4 supergravity coupled to N=4 matter and hidden symmetries}.
\newblock {\em Nucl.Phys.}, {\bf B 185},\ 403--415, 1981.

\bibitem{Duff86}
M.J. Duff, B.E.W. Nillson, and C.N. Pope.
\newblock {\ssl Kaluza--Klein supergravity}.
\newblock {\em Phys.Rep.}, {\bf 130},\ 1--142, 1986.

\bibitem{Freedman83}
D.Z. Freedman, G.W. Gibbons, and P.C. West.
\newblock {\ssl Ten into four won't go}.
\newblock {\em Phys.Lett.}, {\bf B 124},\ 491--492, 1983.

\bibitem{Freedman84}
D.Z. Freedman and G.W. Gibbons.
\newblock {\ssl Electrovac ground state in gauged SU(2)$\times$SU(2)
  supergravity}.
\newblock {\em Nucl.Phys.}, {\bf B 233},\ 24--49, 1984.

\bibitem{Cowdall97}
P.M. Cowdall.
\newblock {\ssl Supersymmetric electrovacs in gauged supergravities}.
\newblock {\em Class.Quant.Grav.}, {\bf 15},\ 2937--2953, 1998.

\bibitem{Singh98}
H.~Singh.
\newblock {\ssl New supersymmetric vacua for D=4, N=4 gauged supergravity}.
\newblock {\em Phys.Lett.}, {\bf B 429},\ 304--312, 1998.

\bibitem{Klemm98}
D.~Klemm.
\newblock {\ssl BPS black holes in gauged N=4, D=4 supergravity}.
\newblock {\em Nucl.Phys.}, {\bf B 545},\ 461--478, 1999.

\bibitem{Cowdall98}
P.M. Cowdall and P.K. Townsend.
\newblock {\ssl Gauged supergravity vacua from intersecting branes}.
\newblock {\em Phys.Lett.}, {\bf B 429},\ 281--288, 1998.

\bibitem{Singh98a}
H.~Singh.
\newblock {\ssl Intersecting branes and anti-de Sitter spacetimes in
  SU(2)$\times$SU(2) gauged supergravity}.
\newblock {\em Phys.Lett.}, {\bf B 444},\ 327--331, 1998.

\bibitem{Manton83}
N.S. Manton.
\newblock {\ssl Topology in the Weinberg-Salam theory}.
\newblock {\em Phys.Rev.}, {\bbf D 28},\ 2019--2026, 1983.

\bibitem{Donets93}
E.E. Donets and D.V. Gal'tsov.
\newblock {\ssl Stringy sphalerons and non-Abelian black holes}.
\newblock {\em Phys.Lett.}, {\bbf B 302},\ 411--418, 1993.

\bibitem{Lavrelashvili93}
G.~Lavrelashvili and D.~Maison.
\newblock {\ssl Regular and black hole solutions of Einstein -- Yang -- Mills
  dilaton theory}.
\newblock {\em Nucl.Phys.}, {\bbf B 410},\ 407--422, 1993.

\bibitem{Bizon93a}
P.~Bizon.
\newblock {\ssl Saddle points of stringy action}.
\newblock {\em Acta Phys.Pol.}, {\bbf B 24},\ 1209--1220, 1993.

\bibitem{Boonstra98}
H.J.~Boonstra, B.~Peeters, and K.~Skenderis. 
\newblock {\ssl Brane intersections, anti-de Sitter spacetimes
and dual superconformal theories}.
\newblock {\em Nucl.Phys.}, {\bbf B 533},\ 127--162, 1998.


\bibitem{Giveon98}
A.~Giveon, D.~Kutasov, and N.~Seiberg. 
\newblock {\ssl Comments on string theory on $AdS_3$}.
\newblock {\em Adv.Theor.Math.Phys.}, {\bbf 2},\ 733--780, 1998.

\bibitem{Behrndt98}
K.~Behrndt, I.~Brunner, and I.~Gaida 
\newblock {\ssl AdS(3) gravity and conformal field theories}.
\newblock {\em Nucl.Phys.}, {\bbf B 546},\ 65--95, 1999.

\end{thebibliography}

\end{document}